\documentclass[aps,pre,preprint,showpacs]{revtex4}
\usepackage{amssymb}

\usepackage{amsmath}
\usepackage{graphicx}

\begin{document}

\title{Hydrodynamics of an inelastic gas with implications for sonochemistry}
\author{James F. Lutsko}
\affiliation{Center for Nonlinear Phenomena and Complex Systems, Universit\'{e} Libre de
Bruxelles, C.P. 231, Blvd. du Triomphe, 1050 Brussels, Belgium}
\email{jlutsko@ulb.ac.be}
\date{\today }
\pacs{47.70.Nd, 47.40.Nm,51.20.+d,78.60.Mq,82.40.Fp}

\begin{abstract}
The hydrodynamics for a gas of hard-spheres which sometimes experience
inelastic collisions resulting in the loss of a fixed, velocity-independent,
amount of energy $\Delta $ is investigated with the goal of understanding
the coupling between hydrodynamics and endothermic chemistry. The
homogeneous cooling state of a uniform system and the modified Navier-Stokes
equations are discussed and explicit expressions given for the pressure,
cooling rates and all transport coefficients for D-dimensions. The
Navier-Stokes equations are solved numerically for the case of a
two-dimensional gas subject to a circular piston so as to illustrate the
effects of the enegy loss on the structure of shocks found in cavitating
bubbles. It is found that the maximal temperature achieved is a sensitive
function of $\Delta $ with a minimum occuring near the physically important
value of $\Delta \sim 12,000K \sim 1eV$.
\end{abstract}

\maketitle

\section{Introduction}

Sonochemistry allows for the achievement of extreme temperatures, pressures
and densities in simple, table-top experiments\cite{SonoChem}. This is
because ultrasound excites bubble cavitation which in turn gives rise to
these extreme conditions. The results are important both because bubble
cavitation occurs naturally\cite{Bubbles} and because of the theoretical and
technological interest in both bubble cavitation and ultrasound technology%
\cite{SonoChem,SonoReview}. One of the most well-known manifestations
of sonochemistry is sonoluminescence whereby, under certain conditions,
fluids irradiated with ultrasound are observed to emit light\cite{SonoReview}%
. The conversion of mechanical energy into electromagnetic form occurs at
the high temperatures and pressures achieved during bubble cavitation,
although many details such as the composition of the gases inside the
bubbles and the relative importance of various light-generating mechanisms
remains unclear. Many recent theoretical studies have focused on modeling
the behavior of a single gas bubble embedded in a liquid and subjected to
sound waves\cite{SonoModelWu,SonoModelMoss,SonoModelKondic,SonoModelVuong,SonoModelYuan,SonoModelAn}. The predicted
temperatures achieved at the points of maximum compression are typically of
the order $10^{4}-10^{5}K$ and shocks are sometimes observed to form
although their presence depends on the exact conditions of the gas and some
details of the modeling. Under these extreme conditions a variety of
activated chemical reactions, see e.g. \cite{SonoEnergyLoss},
collision-induced emissions\cite{SonoCollisionEmission,SonoExcitations}, ionization and electron-ion recombination\cite%
{SonoEmission} and bremsstrahlung\cite{FrommholdBremsstrahlung} are all
expected to occur.

All of the studies cited above which take account of chemistry rely on
equilibrium chemical models. The dynamics of the bubble boundary is
described by the Rayleigh-Plesset equation and the chemistry by equilibrium
rate equations. The bubble is either assumed to have uniform temperature,
density and pressure (called adiabatic models) or the dynamics of the gas
inside the bubble is modeled by the Navier-Stokes equations (in which case,
the rate equations couple to the Navier-Stokes equations via a convective
term). In all cases, the rate constants and transport coefficients used are
based on the assumption that only small deviations from local equilibrium
occur. However, it is not at all clear that the interior of the bubble can
be adequately described by a local equilibrium ensemble, particularly when
there is a significant conversion of mechanical energy (from the sound) into
other, non-mechanical, forms via endothermic chemical reactions or
radiation. And indeed, the experimental work of Didenko and Suslick lead
them to conclude that ''... the temperatures attained in single-bubble
cavitation in liquids with significant vapor pressures will be substantially
limited by endothermic chemical reactions of the polyatomic species inside
the collapsing bubble''\cite{SonoEnergyLossExperiment}. This issue has been
investigated by Yasui using an adiabatic model\cite{SonoEnergyLoss} with a
similar conclusion but that work was challenged by Toegel et al.\cite%
{SonoEnergyLossChallenged} who found that including excluded volume effects
in the equation of state of the gas in an adiabatic model eliminated much of
the effect. Aside from the fact that these studies ignore the gas dynamics
within the bubble, there is reason to question both of these calculations as
experience with granular systems has shown that even a small degree of
inelasticity leads to a rich phenomenology of clustering instabilities and
non-intuitive behavior\cite{GranularPhysicsToday,GranularGases,GranularGasDynamics} due to the inherently nonequilibrium nature of the
system. This naturally leads to the question as to whether it makes any
sense to ignore the nonequilibrium effects undoubtedly present during bubble
cavitation.

Indeed, one could go one step further and ask whether the Navier-Stokes
equations are even applicable in the presence of the large gradients
predicted during cavitation. A recent comparison between the Navier-Stokes
equations for a gas of hard spheres subject to a spherical compression and
computer simulations of the same system did indeed give support to the
adequacy of the Navier-Stokes description\cite{Gaspard}. Given this support,
one must ask whether the Navier-Stokes equations plus rate equations are the
correct hydrodynamic description of such systems , which is properly a
question which must be answered by recourse to kinetic theory. The problem
here is that the only tractable kinetic theory for dense systems, and high
density is an issue during bubble cavitation\cite{SonoChem},\cite%
{SonoEnergyLossChallenged}, is the Enskog theory for hard spheres\cite%
{McLennan}. Fortunately, hard-spheres provide a good first approximation to
the properties of real, interacting systems in all important respects
including transport properties, fluid structure and phase behavior (see,
e.g. \cite{HansenMcdonald}). The kinetic theory for chemically reactive hard
spheres and the coupled hydrodynamics and reaction equations have recently
been formulated and discussed\cite{LutskoJCP} and at least three generic
differences from the ''local-equilibrium'' models noted. First, even for a
single species, the heat flux depends on density gradients as well as
temperature gradients and a new transport coefficient must be introduced.
Second, the reaction equations involve new couplings to the velocity field.
Third, the cooling term - which one might introduce ''by hand'' into the
heat equation to account for endothermic reactions - also involves new
couplings to the velocity field. The first and last effect are well-known
for granular fluids\cite{DuftyGranularTransport} but are not normally
considered in the context of cavitation.

The purpose of the present work is to investigate, within the context of a
minimalist microscopic model, the role of the coupling of energy loss to
hydrodynamics self-consistently, taking into account nonequilibrium effects.
One result will be a richer description of the role of energy loss on the
maximum temperature obtained at the center of a cavitating bubble than given
in previous work \cite{SonoEnergyLoss,SonoEnergyLossChallenged}.

The detailed interaction model will be given in the next Section. It
consists of hard-spheres which lose a fixed quantity of energy $\Delta $
upon collision, provided the rest-frame kinetic energy is sufficient. This
is intended as a toy model which resembles an activated chemical process and
which differs from granular fluids in the physically important sense that
the energy loss is (a) discrete and (b) bounded from below - in granular
fluids, a fixed fraction of the kinetic energy is lost in \emph{all}
collisions. This is not intended to model any particular radiation or
chemical mechanism although it might be considered a crude model of
collision-induced excitations and emission. The goal is then to derive from
the kinetic theory a hydrodynamic description of the fluid using the
Chapman-Enksog method\cite{McLennan,LutskoJCP,LutskoCE},.
Since the Chapman-Enskog method of deriving the hydrodynamic equations is
basically a gradient expansion about the homogeneous fluid, the first
question which must be addressed is the nature of the homogeneous state of
the radiating gas. For an equilibrium hard-sphere fluid, the homogeneous
system has constant temperature and the atomic velocities obey a Maxwell
distribution. When inelastic collisions occur, the system loses energy
continuously and the homogeneous state is not so simple: the temperature is
time dependent and the distribution of velocities is no longer Maxwellian.
The calculation of the cooling rate and the lowest-order corrections to the
Maxwell distribution are the subject of Section II of this paper. In Section
III, the transport properties are calculated and it is shown that they
behave anomalously for temperatures near the energy-loss threshold. These
include the shear and bulk viscosities, the thermal conductivity and a new
transport coefficient describing the transport of heat due to density
gradients. The latter is typical of interactions which do not conserve
energy and plays an important role in the instabilities occurring in
granular fluids. In Section IV, the resulting hydrodynamics is used to study
the effects of the energy loss on a two-dimensional gas confined to a
circular volume with a contracting walls - a circular piston, which sets up
shock waves within the gas. It is found that the maximum temperature
obtained lowers as the energy threshold is lowered until a minimum is
reached at which point the maximum temperature increases with decreasing
threshold. In the final Section, it is argued that the minimum may be of
physical significance.

\section{The homogeneous cooling state}

\subsection{Dynamical Model}

Consider a collection of $N$ hard-spheres in $D$ dimensions having diameter $%
\sigma $ and mass $m$. The position and velocity of the i-th particle will
be denoted $\overrightarrow{q}_{i}$ and $\overrightarrow{v}_{i}$,
respectively. The particles are confined to a box of volume $V$ giving a
number density $n=N/V$. The boundary conditions are not important here and
could be, e.g., hard, elastic walls or periodic. The only interactions are
instantaneous collisions:\ between collisions the particles stream freely.
When two particles collide, their positions are unchanged but they lose a
quantity of energy, $\Delta _{a}$ which could be zero. In ref. \cite%
{LutskoJCP} it is shown that in this case, the velocities after the
collision, $\overrightarrow{v}_{i}^{\prime }$ and $\overrightarrow{v}%
_{j}^{\prime }$, will be related to the pre-collisional velocities, $%
\overrightarrow{v}_{i}$ and $\overrightarrow{v}_{j}$, by%
\begin{eqnarray}
\overrightarrow{v}_{i}^{\prime } &=&\overrightarrow{v}_{i}-\frac{1}{2}%
\widehat{q}_{ij}\left( \overrightarrow{v}_{ij}\cdot \widehat{q}%
_{ij}+sgn\left( \overrightarrow{v}_{ij}\cdot \widehat{q}_{ij}\right) \sqrt{%
\left( \overrightarrow{v}_{ij}\cdot \widehat{q}_{ij}\right) ^{2}-\frac{4}{m}%
\Delta _{a}}\right) \\
\overrightarrow{v}_{j}^{\prime } &=&\overrightarrow{v}_{j}+\frac{1}{2}%
\widehat{q}_{ij}\left( \overrightarrow{v}_{ij}\cdot \widehat{q}%
_{ij}+sgn\left( \overrightarrow{v}_{ij}\cdot \widehat{q}_{ij}\right) \sqrt{%
\left( \overrightarrow{v}_{ij}\cdot \widehat{q}_{ij}\right) ^{2}-\frac{4}{m}%
\Delta _{a}}\right)  \notag
\end{eqnarray}%
where $\widehat{q}_{ij}=\left( \overrightarrow{q}_{i}-\overrightarrow{q}%
_{j}\right) /\left| \overrightarrow{q}_{i}-\overrightarrow{q}_{j}\right| $
is the unit vector pointing from the center of the i-th particle to the
center of the j-th particle. This collision rule preserves the total
momentum $m\left( \overrightarrow{v}_{i}+\overrightarrow{v}_{j}\right) $ as
well as the total angular momentum of the colliding particles but allows for
an energy loss as can be seen by computing%
\begin{equation}
\frac{m}{2}v_{i}^{\prime 2}+\frac{m}{2}v_{j}^{\prime 2}=\frac{m}{2}v_{i}^{2}+%
\frac{m}{2}v_{j}^{2}-\Delta _{a}
\end{equation}%
so that $\Delta _{a}$ characterizes the energy loss which may in general be
any function of the normal component of the relative momentum of the
colliding atoms (i.e., $\Delta _{a}=\Delta _{a}(\overrightarrow{v}_{ij}\cdot 
\widehat{q}_{ij})$). The subscript allows for the possibility that different
types of collisions are possible and each collision realizes the a-th
collision type with probability $K_{a}$. To be specific, we will consider
the simplest model of a sudden energy loss wherein for collisions with
enough energy in the normal component of the momenta, i.e. $\frac{m}{4}%
\left( \overrightarrow{v}_{ij}\cdot \widehat{q}_{ij}\right) ^{2}>\Delta $,
an inelastic collision, removing energy $\Delta $ occurs with fixed
probability $p$ and otherwise the collision is elastic. In terms of the step
function $\Theta \left( x\right) $ which is $0$ for $x<1$ and $1$ for $x>1$
, this may be written as 
\begin{eqnarray}
K_{0} &=&\Theta \left( \Delta -\frac{m}{4}\left( \overrightarrow{v}%
_{ij}\cdot \widehat{q}_{ij}\right) ^{2}\right) +\left( 1-p\right) \Theta
\left( \frac{m}{4}\left( \overrightarrow{v}_{ij}\cdot \widehat{q}%
_{ij}\right) ^{2}-\Delta \right) ,\;\Delta _{0}=0 \\
K_{1} &=&p\Theta \left( \frac{m}{4}\left( \overrightarrow{v}_{ij}\cdot 
\widehat{q}_{ij}\right) ^{2}-\Delta \right) ,\;\Delta _{1}=\Delta  \notag
\end{eqnarray}%
where the $a=0$ collisions are elastic and the $a=1$ collisions inelastic.
For $p=0$ the collisions are purely elastic and for $p=1$ the collisions are
always inelastic if enough energy is available.

\subsection{General description of the homogeneous cooling state}

Given this model, for $p>0$ the system will always cool since even for very
low temperatures, there will be some population of sufficiently energetic
atoms which undergo inelastic collisions. Thus, the simplest state the
system can experience is one which is spatially homogeneous but undergoing
continual cooling, the so-called Homogeneous Cooling State (HCS)\ which is
the analog of the equilibrium state for this intrinsically non-equilibrium
system. In this case, the velocity distribution will not be strictly
Gaussian. In fact, it has been shown\cite{LutskoCE} that the nonequilibrium
velocity distribution can be written as an expansion about a simple Gaussian
distribution so that 
\begin{equation}
f_{0}(\overrightarrow{v};T\left( t\right) )=f_{M}\left( \overrightarrow{v}%
;T(t)\right) \sum_{k}c_{k}\left( T(t)\right) L_{k}^{\frac{D-2}{2}}\left( 
\frac{m}{2k_{B}T\left( t\right) }v^{2}\right)  \label{hsc-distribution}
\end{equation}%
where the Maxwellian distribution for a fluid with uniform number density $n$%
and temperature $T$ is 
\begin{equation}
f_{M}\left( \overrightarrow{v};T\left( t\right) \right) =n\pi ^{-D/2}\left( 
\frac{2k_{B}T\left( t\right) }{m}\right) ^{-D/2}\exp \left( -\frac{m}{%
2k_{B}T\left( t\right) }v_{1}^{2}\right)
\end{equation}%
and the associated Laguerre polynomials are 
\begin{equation}
L_{k}^{\frac{D-2}{2}}\left( x\right) =\sum_{m=0}^{k}\frac{\Gamma \left( 
\frac{D}{2}+k\right) \left( -x\right) ^{m}}{\Gamma \left( \frac{D}{2}%
+m\right) \left( k-m\right) !m!}.
\end{equation}%
The temperature cools according to 
\begin{equation}
\frac{d}{dt}T\left( t\right) =\frac{2}{Dnk_{B}}\xi ^{(0)}  \label{cool1}
\end{equation}%
with a cooling rate given by%
\begin{equation}
\frac{2}{Dnk_{B}}\xi ^{(0)}=-T\sum_{rs}I_{1,rs}c_{r}c_{s}.  \label{cooling}
\end{equation}%
The first two coefficients in the expansion of the distribution are $c_{0}=1$
and $c_{1}=0$ which follow from the definition of the density and temperature%
\cite{LutskoJCP}. The higher order coefficients obey%
\begin{equation}
\left( \frac{2}{Dnk_{B}T}\xi ^{(0)}\right) \left[ T\frac{\partial }{\partial
T}c_{k}+k\left( c_{k}-c_{k-1}\right) \right] =\sum_{rs}I_{k,rs}c_{r}c_{s}
\label{a0}
\end{equation}%
where the coefficients $I_{k.rs}$ appearing in these equations can be
calculated from a generating function\cite{LutskoCE}, details of which can
be found in Appendix \ref{AppA}. It is convenient to use dimensionless
variables $\Delta ^{\ast }=\Delta /k_{B}T$, $n^{\ast }=n\sigma ^{D}$, 
\begin{equation}
t^{\ast }=2p\left( D+2\right) n^{\ast }\chi \frac{S_{D}}{2D\left( D+2\right) 
\sqrt{\pi }}\left( \frac{\Delta }{m\sigma ^{2}}\right) ^{1/2}t
\end{equation}%
and 
\begin{equation}
I_{k,rs}=n^{\ast }\chi \frac{S_{D}}{2D\left( D+2\right) \sqrt{\pi }}\left( 
\frac{k_{B}T}{m\sigma ^{2}}\right) ^{1/2}I_{k,rs}^{\ast },
\end{equation}%
where the area of the unit $D$-sphere is 
\begin{equation}
S_{D}=\frac{2\pi ^{D/2}}{\Gamma \left( D/2\right) }.
\end{equation}%
The quantity $\chi $ occurring in these expressions is the pair distribution
function at contact which occurs in the Enskog theory to take account of the
enhancement of the collision rate due to the finite size of the atoms. In
the low density (Boltzmann) limit, it goes to one. It is solely a function
of density, and at finite densities, it is usually approximated as that of a
local equilibrium fluid of hard spheres. When performing numerical
calculations, the value used for two-dimensional disks will be that of the
approximation due to Henderson see e.g. \cite{BarkerHend_WhatIsLiquid}, 
\begin{equation}
\chi =\frac{1-\frac{7y}{16}}{(1-y)^{2}}
\end{equation}%
with $y=\pi n\sigma ^{2}/4$ while for three dimensional spheres the
Carnahan-Starling expression\cite{BarkerHend_WhatIsLiquid} 
\begin{equation}
\chi =\frac{1-\frac{y}{2}}{(1-y)^{3}}
\end{equation}%
with $y=\pi n\sigma ^{3}/6$ will be used.

The evolution of the temperature is then given by%
\begin{equation}
\frac{d}{dt}T\left( t\right) =-n^{\ast }\chi \frac{S_{D}}{2D\left(
D+2\right) \sqrt{\pi }}\left( \frac{k_{B}T}{m\sigma ^{2}}\right)
^{1/2}T\sum_{rs}I_{1,rs}^{\ast }c_{r}c_{s}
\end{equation}%
or 
\begin{equation}
\frac{d}{dt^{\ast }}\Delta ^{\ast }=\frac{\left( \Delta ^{\ast }\right)
^{1/2}}{2p\left( D+2\right) }\sum_{rs}I_{1,rs}^{\ast }c_{r}c_{s}
\label{coolinga}
\end{equation}%
and the coefficients of the expansion of the distribution function are%
\begin{equation}
\frac{d}{dt^{\ast }}c_{k}=\frac{1}{2p\left( D+2\right) \left( \Delta ^{\ast
}\right) ^{1/2}}\left[ \sum_{rs}I_{k,rs}^{\ast }c_{r}c_{s}+\left(
\sum_{rs}I_{1,rs}^{\ast }c_{r}c_{s}\right) k\left( c_{k}-c_{k-1}\right) %
\right] ,  \label{ck}
\end{equation}%
where eq.(\ref{coolinga}) has been used to replace $\frac{d}{d\Delta ^{\ast }%
}c_{k}$ by $\frac{d}{dt^{\ast }}c_{k}$. All of these expressions are exact
consequences of the Enskog kinetic theory\cite{LutskoCE}. Generally, three
approximations are made in order to solve these equations:\ (1)one sets $%
c_{k}=0$ for $k>k_{0}$ for some integer $k_{0};$ (2) only the first $k_{0}$
equations of the coupled hierarchy given in eq.(\ref{a0}) are retained and
(3) one also neglects quadratic terms $c_{r}c_{s}$ for $r+s>k_{0}$ and cubic
terms $c_{r}c_{s}c_{k}$ for $r+s+k>k_{0}$. The first two approximations are
necessary in order to make solution of the equations possible while the
third is more a matter of convenience which generally introduces no
significant errors.

In the following Subsections, the lowest two orders of approximation will be
evaluated. In order to judge the accuracy of these approximations, the
results will be compared to the numerical solution of the Enskog equation by
means of the Direct Simulation Monte Carlo (DSMC) method formulated by Bird%
\cite{BirdDSMC} for the Boltzmann equation and extended to the Enskog
equation by Santos et al\cite{SantosDSMC}. The results given below were
obtained from runs in three dimensions using $10^{5}$ points in a cubic
volume of $\sigma ^{3}$ with periodic boundaries. The timestep used was
0.0117 mean free times and the length of the run was 2000 mean free times.

\subsection{The Local Equilibrium Approximation}

\begin{figure}[tbp]
\begin{center}
\resizebox{17.4cm}{!}{
{\includegraphics{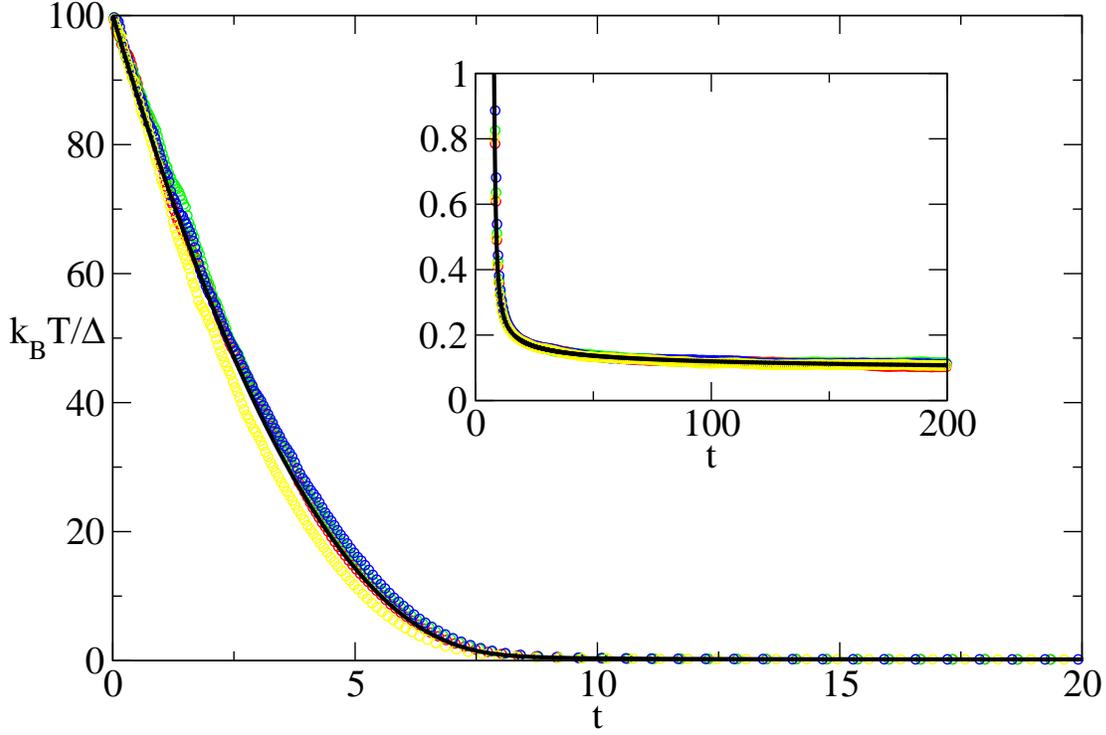}}}
\end{center}
\caption{The temperature as a function of time as predicted by eq. \ref%
{temp1} (line) and as determined from four different DSMC simulations
(circles of different colors). The theory and simulation results are
indistinguishable. Note the very rapid, algebraic, decrease in the
temperature at early times when the dimensionless temperature is large and
the very slow, logarithmic, decay at long times when the dimensionless
temperature is less than one.}
\label{fig1}
\end{figure}

\begin{figure}[tbp]
\begin{center}
\resizebox{17.4cm}{!}{
{\includegraphics{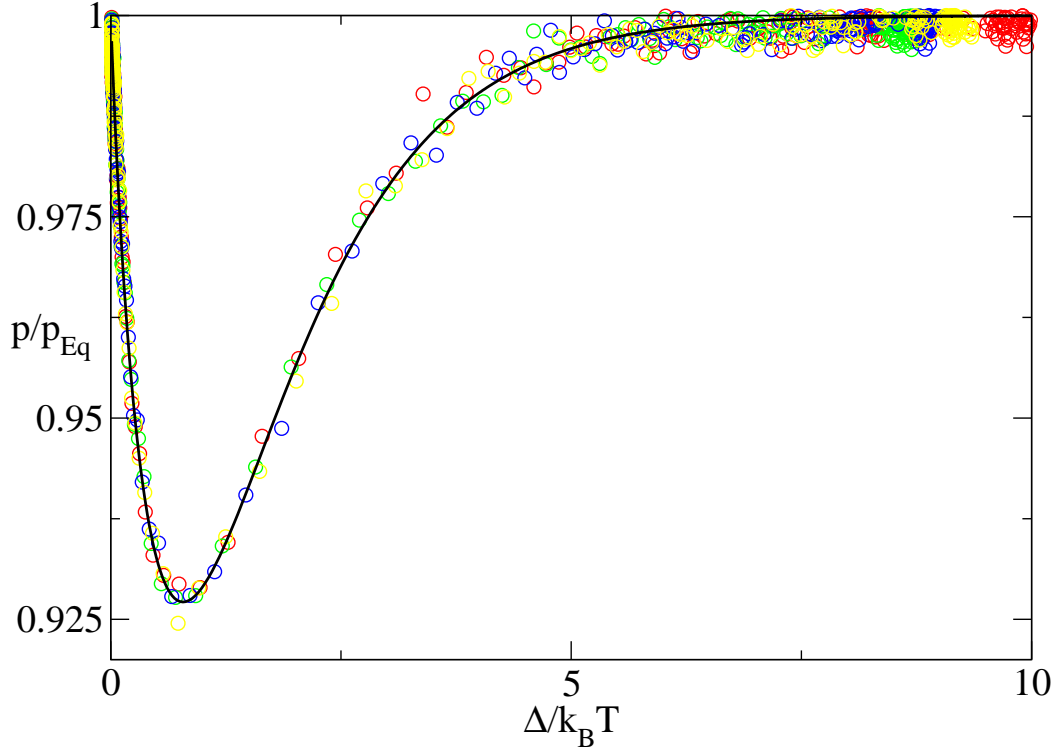}}}
\end{center}
\caption{The ratio of the pressure of the dissipitive gas to that of an
equilbrium $\Delta=0$ gas as a function of $\Delta^{*}$ for a dense ($%
n^{*}=0.5$) gas in three dimensions with $p=1$.}
\label{fig2}
\end{figure}

The simplest approximation is to take $k_{0}=1$ which thus ignores all
non-equilibrium corrections to the distribution. Then, as discussed in
Appendix \ref{AppA}, one has that%
\begin{equation}
I_{1,00}^{\ast }=2p\left( D+2\right) \Delta ^{\ast }e^{-\Delta ^{\ast }},
\label{I100}
\end{equation}%
and eq.(\ref{cool1}) becomes 
\begin{equation}
\frac{d}{dt^{\ast }}\Delta ^{\ast }=\Delta ^{\ast 3/2}e^{-\Delta ^{\ast }}.
\label{temp1}
\end{equation}%
Thus, at this level of approximation, all details of the probability of an
inelastic collision, the density and the dimensionality can be scaled out of
the expression for the temperature so that the temperature follows a
universal curve. As expected, eq.(\ref{temp1}) gives a non-zero cooling rate
for all values of the temperature. Note that right hand side goes to zero in
the two limits $\Delta ^{\ast }\rightarrow 0$ and $\Delta ^{\ast
}\rightarrow \infty $. This reflects the fact that at very low temperatures, 
$\Delta ^{\ast }\rightarrow \infty $, very few collisions occur with enough
energy to be inelastic while at very high temperature, $\Delta ^{\ast
}\rightarrow 0$, the loss of energy during inelastic collisions is of no
physical importance since it represents a vanishing fraction of the total
energy of each atom. Both of these limits can therefore be viewed, in some
sense, as ''elastic'' limits although in neither case can one say that only
elastic collisions occur and it is to be expected that in both of these
''elastic'' limits, all thermodynamic and transport properties will be those
of an elastic gas since the energy dissipation plays no role. The source
term in eq.(\ref{temp1}) also exhibits a maximum at $\Delta ^{\ast }=3/2$
which is not surprising and in fact must be true of the exact cooling rate:
since $\Delta ^{\ast }$ goes to zero in the elastic limits it must either be
constant or it must have at least one extremum. Nevertheless, when written
in terms of dimensional quantities, the cooling equation becomes 
\begin{equation}
\frac{d}{dt}T=-2p\left( D+2\right) n^{\ast }\chi \frac{S_{D}}{2D\left(
D+2\right) \sqrt{\pi }}\left( \frac{\Delta }{m\sigma ^{2}}\right) ^{1/2}%
\sqrt{\Delta T}e^{-\Delta /T}
\end{equation}%
showing that the physical cooling rate is a monotonically increasing
function of the temperature. For high temperatures, $\Delta ^{\ast }=\beta
\Delta \ll 1$ and has the approximate behavior 
\begin{equation}
\Delta ^{\ast }\left( t^{\ast }\right) \simeq \frac{\Delta ^{\ast }(0)}{%
\left( 1-\frac{1}{2}\sqrt{\Delta ^{\ast }(0)}t^{\ast }\right) ^{2}}
\end{equation}%
so that the temperature decays algebraically. When the temperature is low, $%
\Delta ^{\ast }=\beta \Delta \gg 1$and it can be shown (see Appendix
\ref{AppB}) that the cooling becomes very slow ($\Delta ^{\ast }\sim \ln
t^{\ast }$). The full behavior is shown in Fig. \ref{fig1} where the
crossover between these two asymptotic forms is evident. As the figure
shows, the theory is in good agreement with the simulations. In this
approximation, the pressure is given by%
\begin{equation} \label{pressure}
\frac{p}{nk_{B}T}=1+n^{\ast }\chi \frac{S_{D}}{2D}+n^{\ast }\chi \frac{S_{D}%
}{4D}p\left( \left( 1-2\sqrt{\frac{\Delta ^{\ast }}{\pi }}\right) e^{-\Delta
^{\ast }}+\left( \text{erf}\left( \sqrt{\Delta ^{\ast }}\right) -1\right)
\right) .
\end{equation}%
The first two terms on the right are the usual equilibrium expression and
the third terms represents the non-equilibrium correction. The effect of the
nonequilibrium term is shown in Fig. \ref{fig2} where it is seen that a
minimum occurs in $p/nkT$ near $\Delta /k_{B}T=1$. The effect is small, the
minimum being about 8\% below the equilibrium value. Again, the theory and
simulation are seen to be in good agreement.

\subsection{First non-Gaussian correction}

\begin{figure}[tbp]
\begin{center}
\resizebox{17.4cm}{!}{
{\includegraphics{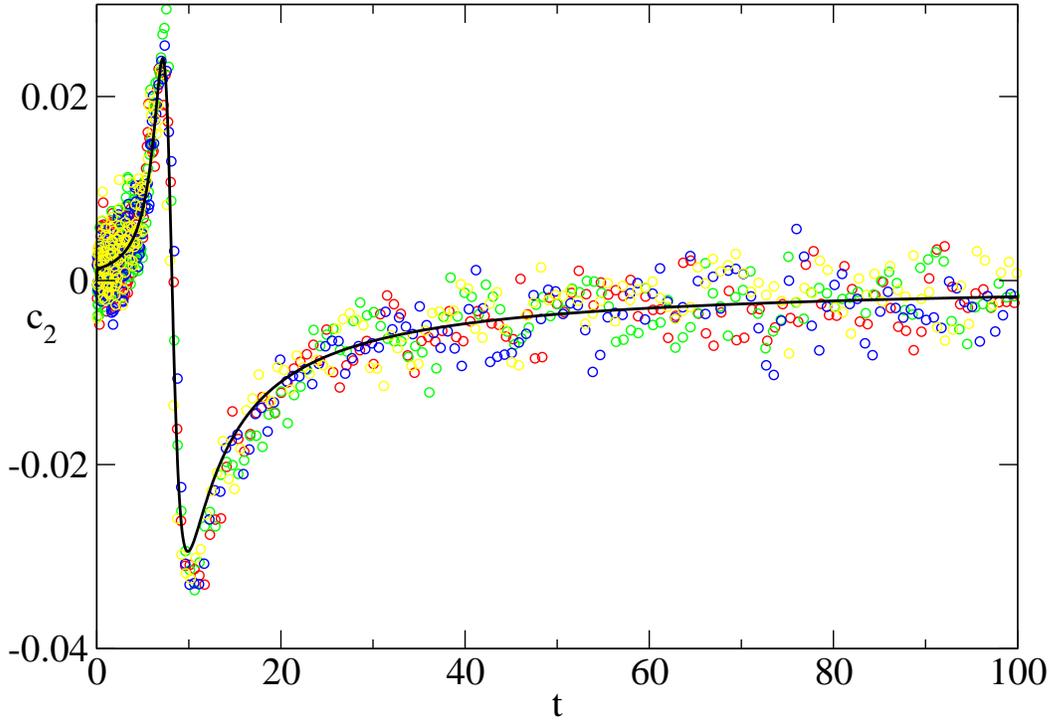}}}
\end{center}
\caption{The coefficient of the first non-Gaussian correction, $c_{2}$ as
predicted by eq. \ref{ck} (line) and as determined from four different DSMC
simulations (circles). The insert shows the rapid convergence of the
numerical solution for three different initial conditions.}
\label{fig3}
\end{figure}

The simplest non-gaussian correction consists of taking $k_{0}=2.$The
required coefficients are, in addition to $I_{1,00}^{\ast }$ given above, 
\begin{eqnarray}
I_{2,00}^{\ast } &=&-2p\Delta ^{\ast }e^{-\Delta ^{\ast }}\left( \Delta
^{\ast }-1\right) \\
I_{1,02}^{\ast }+I_{1,20}^{\ast } &=&p\frac{D+2}{8}\Delta ^{\ast }e^{-\Delta
^{\ast }}\left( 4\Delta ^{\ast }{}^{2}-4\Delta ^{\ast }-1\right)  \notag \\
I_{2,02}^{\ast }+I_{2,20}^{\ast } &=&-8\left( D-1\right) -\frac{1}{8}%
pe^{-\Delta ^{\ast }}\left( 4\Delta ^{\ast }{}^{4}-8\Delta ^{\ast
}{}^{3}+\left( 75+16D\right) \Delta ^{\ast 2}+\left( 69-24D\right) \Delta
^{\ast }+32\left( 1-D\right) \right)  \notag \\
&&-2p\left( D-1\right) \Delta ^{\ast }e^{-\frac{1}{2}\Delta ^{\ast
}}K_{1}\left( \frac{1}{2}\Delta ^{\ast }\right)  \notag
\end{eqnarray}%
where $K_{\nu }\left( x\right) $ is the modified Bessel function of the
second kind. Equations (\ref{coolinga}) and (\ref{ck}) can be solved
simultaneously given an initial temperature $\Delta ^{\ast }\left( 0\right) $
and a value for $c_{2}\left( t=0\right) =c_{2}\left( \Delta ^{\ast }=\Delta
^{\ast }\left( 0\right) \right) $. Figure \ref{fig3} shows $c_{2}\left(
t^{\ast }\right) $ for $n^{\ast }=0.5$ in three dimensions as well as the
results obtained from the simulation. The behavior of $c_{2}$ is seen to be
highly non-trivial with a sharp change of sign at $\Delta ^{\ast }~=1$ and,
for longer times, a decay to zero. The inset shows that the initial
condition for $c_{2}$ is irrelevant except at very short times. Overall, the
agreement between the calculation and the simulation is very good. The small
magnitude of $c_{2}$ indicates that the distribution can be well
approximated by a Gaussian and the agreement of the pressure to the
simulations support the view that the homogeneous state is well-described by
a local-equilibrium distribution.

\section{Transport Coefficients}

The Navier-Stokes equations for a dissipative system take the form\cite%
{LutskoJCP} 
\begin{eqnarray}
\frac{\partial }{\partial t}n+\overrightarrow{\nabla }\cdot \left( 
\overrightarrow{u}n\right) &=&0 \label{NS} \\ 
\frac{\partial }{\partial t}\overrightarrow{u}+\overrightarrow{u}\cdot 
\overrightarrow{\nabla }\overrightarrow{u}+\frac{1}{\rho }\overrightarrow{%
\nabla }\cdot \overleftrightarrow{P} &=&0  \notag \\
\left( \frac{\partial }{\partial t}+\overrightarrow{u}\cdot \overrightarrow{%
\nabla }\right) T+\frac{2}{Dnk_{B}}\left[ \overleftrightarrow{P}:%
\overrightarrow{\nabla }\overrightarrow{u}+\overrightarrow{\nabla }\cdot 
\overrightarrow{q}\right] &=&\xi _{0}+\xi _{1}\overrightarrow{\nabla }\cdot 
\overrightarrow{u}  \notag
\end{eqnarray}%
with pressure tensor%
\begin{equation} \label{PT}
P_{ij}=p^{(0)}\delta _{ij}-\eta \left( \partial _{i}u_{j}+\partial _{j}u_{i}-%
\frac{2}{D}\delta _{ij}\left( \overrightarrow{\nabla }\cdot \overrightarrow{u%
}\right) \right) -\gamma \delta _{ij}\left( \overrightarrow{\nabla }\cdot 
\overrightarrow{u}\right)
\end{equation}%
and heat-flux vector%
\begin{equation} \label {HFV}
\overrightarrow{q}\left( \overrightarrow{r},t\right) =-\mu \overrightarrow{%
\nabla }\rho -\kappa \overrightarrow{\nabla }T
\end{equation}%
There are two differences between these equations and the Navier-Stokes
equations for an elastic fluid. First, the temperature equation contains
source terms that account for the collisional cooling. Second, the heat flux
vector depends on gradients in density as well as temperature. Both of these
contributions are well-known from the study of granular media. In this
Section, the expressions for the various transport coefficients and source
terms are discussed for the dissipative interaction model.

For equilibrium fluids, the transport coefficients are algebraic functions
of the temperature and density. This is no longer true in the present case
and the transport coefficients must be determined by solving ordinary
differential equations as was also the case with $c_{2}$ above. In the
following, we evaluate the expressions for the transport coefficients given
in ref.\cite{LutskoCE} in the approximation that $c_{2}=0$. Then, from the
previous Section, one has that%
\begin{equation}
\xi _{0}=-\frac{Dnk_{B}T}{2}n^{\ast }\chi \frac{S_{D}}{2D\left( D+2\right) 
\sqrt{\pi }}\left( \frac{k_{B}T}{m\sigma ^{2}}\right) ^{1/2}I_{1,00}^{\ast }
\end{equation}%
It is convenient to express the transport coefficients in terms of
dimensionless functions by writing 
\begin{eqnarray}
\eta &=&\eta _{0}\overline{\eta } \\
\gamma &=&\eta _{0}\overline{\gamma }  \notag \\
\mu &=&\kappa _{0}\frac{T}{n}\overline{\mu }  \notag \\
\kappa &=&\kappa _{0}\overline{\kappa }  \notag
\end{eqnarray}%
where the Boltzmann transport coefficients for an elastic gas are 
\begin{eqnarray}
\eta _{0} &=&\sqrt{mk_{B}T}\frac{\left( D+2\right) \sqrt{\pi }}{4S_{D}}%
\sigma ^{1-D} \\
\kappa _{0} &=&k_{B}\sqrt{\frac{k_{B}T}{m}}\frac{D\left( D+2\right) ^{2}%
\sqrt{\pi }}{8S_{D}\left( D-1\right) }\sigma ^{1-D}.  \notag
\end{eqnarray}%
Then, the transport coefficients can be written in terms of their kinematic
contributions as%
\begin{eqnarray}
\overline{\eta } &=&\left( 1+\frac{2}{3}\overline{\theta }\right) \overline{%
\eta }^{K}+\frac{D}{D+2}\overline{\gamma }_{1} \\
\overline{\gamma } &=&\overline{\gamma }^{K}+\overline{\gamma }_{1}  \notag
\\
\overline{\mu } &=&\left( 1+\overline{\theta }\right) \overline{\mu }^{K} 
\notag \\
\overline{\kappa } &=&\left( 1+\overline{\theta }\right) \overline{\kappa }%
^{K}+\frac{D-1}{D+2}\overline{\gamma }_{1}  \notag
\end{eqnarray}%
where the auxiliary functions $\overline{\gamma }_{1}\left( n^{\ast },\Delta
^{\ast }\right) $ and $\overline{\theta }\left( n^{\ast },\Delta ^{\ast
}\right) $ are given in Appendix \ref{App:Transport}.

As shown in Appendix \ref{App:Transport}, the kinematic contributions to the
transport coefficients satisfy the equations%
\begin{gather}
I_{1,00}^{\ast }\Delta ^{\ast }\frac{\partial \overline{\eta }^{K}}{\partial
\Delta ^{\ast }}+\left( 8DI_{\eta }^{\ast }-\frac{1}{2}I_{1,00}^{\ast
}\right) \overline{\eta }^{K}=\frac{8D}{\chi }+8n^{\ast }\frac{S_{D}}{D+2}%
\Omega _{\eta }^{\ast }  \label{t1} \\
I_{1,00}^{\ast }\Delta ^{\ast }\frac{\partial }{\partial \Delta ^{\ast }}%
\overline{\gamma }^{K}+\left( 8\left( D-1\right) I_{\lambda }^{\ast }-\frac{3%
}{2}I_{1,00}^{\ast }\right) \overline{\gamma }^{K}=16\sqrt{\pi }\Omega
_{\gamma }^{\ast }  \notag \\
I_{1,00}^{\ast }\Delta ^{\ast }\frac{\partial \overline{\kappa }^{K}}{%
\partial \Delta ^{\ast }}+\left( 8\left( D-1\right) I_{\kappa }^{\ast
}-\left( 1+\Delta ^{\ast }\right) I_{1,00}^{\ast }\right) \overline{\kappa }%
^{K}=8\frac{D-1}{\chi }+\frac{12\left( D-1\right) }{D\left( D+2\right) }%
n^{\ast }S_{D}\Omega _{\kappa }^{\ast }  \notag \\
\allowbreak I_{1,00}^{\ast }\Delta ^{\ast }\frac{\partial \overline{\mu }^{K}%
}{\partial \Delta ^{\ast }}+\left( 8\left( D-1\right) I_{\mu }^{\ast }-\frac{%
3}{2}\allowbreak I_{1,00}^{\ast }\right) \overline{\mu }^{K}+I_{1,00}^{\ast
}\left( 1+n\frac{\partial \ln \chi _{0}}{\partial n}\right) \overline{\kappa 
}^{K}=\Omega _{\mu }^{\ast }  \notag
\end{gather}%
where explicit expressions for the Boltzmann integrals, $I_{\alpha }^{\ast }$%
, and the sources, $\Omega _{\alpha }^{\ast }$, are given in Appendix \ref%
{App:Transport}. For both an elastic hard-sphere gas as well as a simple
granular fluid, the dimensionless transport coefficients are only functions
of the density so that the derivative terms vanish and these reduce to a
system of simple algebraic equations. Here, the presence of an additional
energy scale leads to a more complex dependence on temperature.

\begin{figure}[tbp]
\begin{center}
\resizebox{17.4cm}{!}{
{\includegraphics{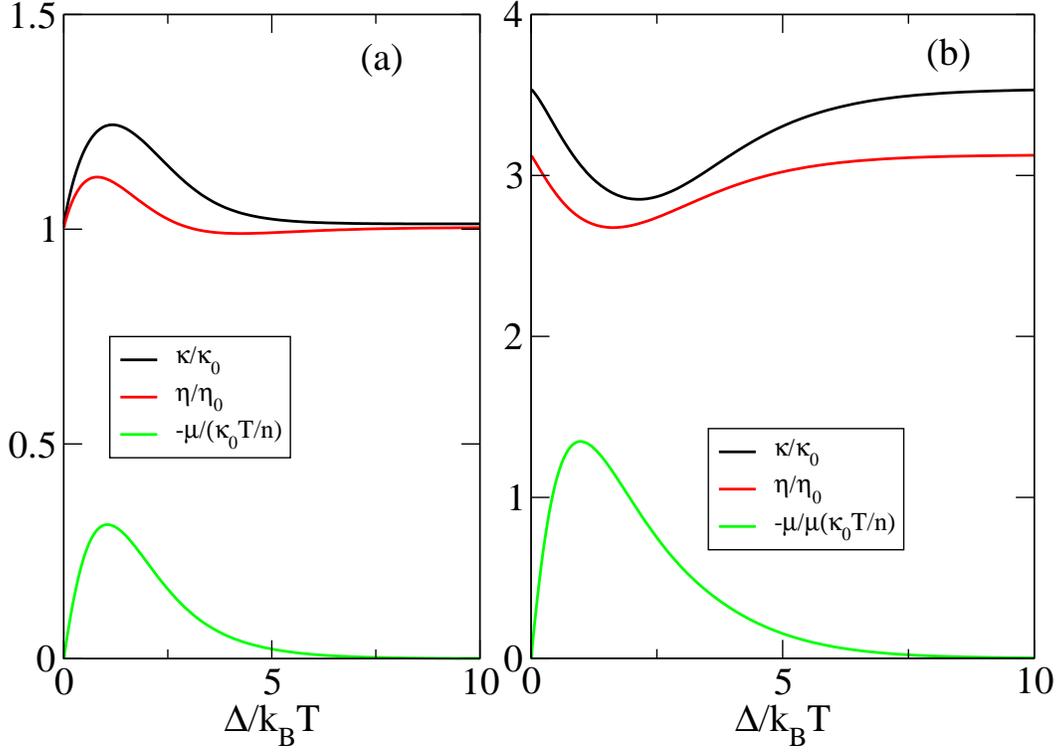}}}
\end{center}
\caption{The dimensionless transport coefficients as funtions of the reduced
temperature for a three dimensional gas with $p=1$ and for (a) a low
density, $n^{*}=0.01$, gas and (b) a high density, $n^{*}=0.5$, gas. The
upper line is the thermal conductivity, the middle line is the shear
viscosity and the bottom line is $-\protect\mu$.}
\label{fig4}
\end{figure}

Figure \ref{fig4} shows the transport coefficients for a low density, $%
n^{\ast }=0.01$, and a moderately dense density gas, $n^{\ast }=0.5,$ with $%
p=1$ in three dimensions as a function of temperature as obtained by solving
these equations with the initial condition that the scaled transport
coefficients take on their equilibrium values at $\Delta ^{\ast }=0$. The
behavior of the transport coefficients is non-monotonic with a ''resonance''
occurring in the vicinity of $\Delta ^{\ast }=1$. For the thermal
conductivity and the shear viscosity, the effect is a reduction of about
20\% compared to the equilibrium transport coefficients while the effect on
the bulk viscosity is minimal. The new transport coefficient $\mu $ is
surprisingly large, being of order 1 up to $3\succsim \Delta ^{\ast
}\succsim 0.05$ and is of order 0.1 for $5\succsim \Delta ^{\ast }\succsim
0.01$.

\begin{figure}[tbp]
\begin{center}
\resizebox{17.4cm}{!}{
{\includegraphics{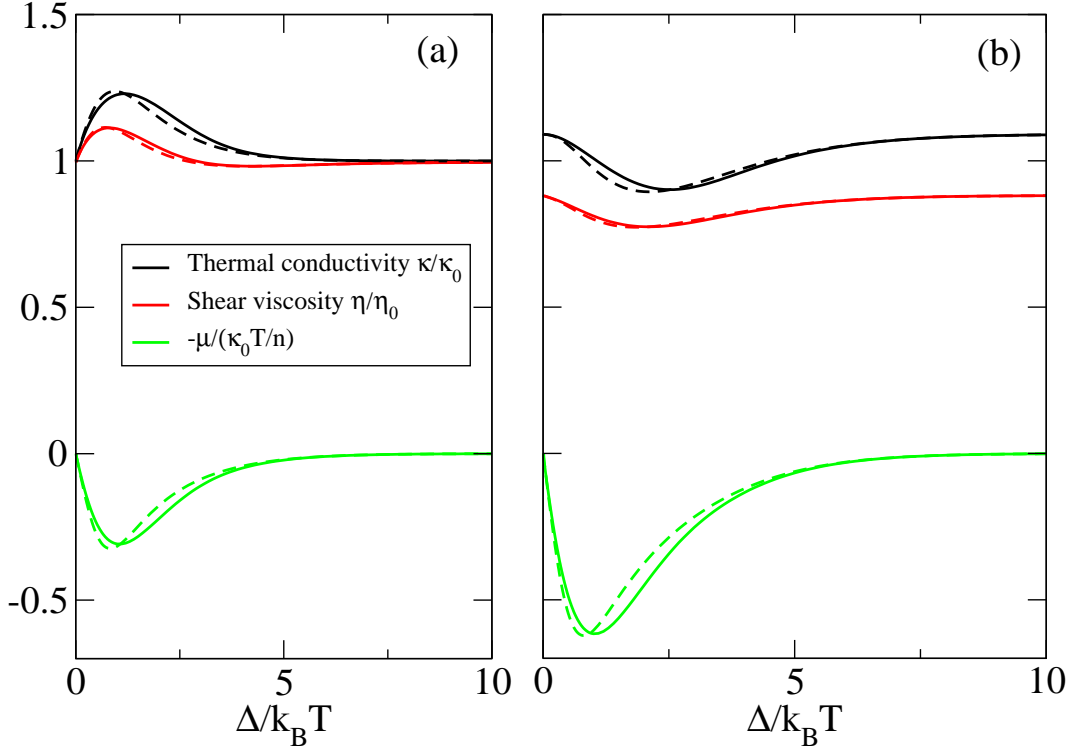}}}
\end{center}
\caption{The dimensionless kinetic contributions to the transport
coefficients as funtions of the reduced temperature.The broken lines are
from the adiabatic approximation.The upper lines are the thermal
conductivity, the middle lines are the shear viscosity and the bottom lines
are $-\protect\mu$.}
\label{fig5}
\end{figure}

Since in both the elastic and the simple granular fluids\cite{LutskoCE}, the
scaled transport coefficients do not depend on temperature, a simpler and
more practical approximation that suggests itself is to drop the derivatives
in eqs.(\ref{t1}) giving%
\begin{gather}
\left( 8DI_{\eta }^{\ast }-\frac{1}{2}I_{1,00}^{\ast }\right) \overline{\eta 
}^{K}=\frac{8D}{\chi }+8n^{\ast }\frac{S_{D}}{D+2}\Omega _{\eta }^{\ast }
\label{t2} \\
\left( 8\left( D-1\right) I_{\lambda }^{\ast }-\frac{3}{2}I_{1,00}^{\ast
}\right) \overline{\gamma }^{K}=16\sqrt{\pi }\Omega _{\gamma }^{\ast } 
\notag \\
\left( 8\left( D-1\right) I_{\kappa }^{\ast }-\left( 1+\Delta ^{\ast
}\right) I_{1,00}^{\ast }\right) \overline{\kappa }^{K}=8\frac{D-1}{\chi }+%
\frac{12\left( D-1\right) }{D\left( D+2\right) }n^{\ast }S_{D}\Omega
_{\kappa }^{\ast }  \notag \\
\left( 8\left( D-1\right) I_{\mu }^{\ast }-\frac{3}{2}\allowbreak
I_{1,00}^{\ast }\right) \overline{\mu }^{K}+I_{1,00}^{\ast }\left( 1+n\frac{%
\partial \ln \chi _{0}}{\partial n}\right) \overline{\kappa }^{K}=\Omega
_{\mu }^{\ast }  \notag
\end{gather}%
which will be termed the adiabatic approximation. Figure \ref{fig5} shows
that eq.(\ref{t2}) provides a reasonable approximation to the full results
obtained by integrating eqs.(\ref{t1}). The reason for the resonance in the
transport coefficients is now easier to see: the Boltzmann integrals $%
I_{\eta }^{\ast },...$ are scaled so as to go to one in the elastic limits $%
\Delta ^{\ast }\rightarrow 0$ and $\Delta ^{\ast }\rightarrow \infty $ when
the energy loss is intelligible and necessarily exhibit at least one
extremum in between. The interaction of these functions with the cooling
term $I_{1,00}^{\ast }$ which behaves similarly with a maximum at $\Delta
^{\ast }=1$, gives rise to the resonances in the transport coefficients
reflecting the fact that the effect of the dissipation is at its maximum
somewhere in between those limits.

\section{Application:\ the circular piston}

In this Section, to illustrate the importance of the coupling of energy loss
and hydrodynamics, the behavior of a gas confined to a contracting circular
piston, described in ref. \cite{Gaspard}, is re-examined for the case of the
model gas considered here. To this end, the modified Navier-Stokes equations, eqs.(\ref{NS})-(\ref{HFV}),
have been solved numerically with a hard circular boundary moving at
constant speed $c$ under the assumption of circular symmetry. The gas is initially in a uniform state at temperature $%
T(0)$ and here the focus will be on the particular case examined in ref. %
\cite{Gaspard} that the wall moves with speed $c=-5\sqrt{\frac{k_{B}T(0)}{m}}
$ and initial number density $n^{\ast }(0)=0.1$. The energy gap parameter
will be reported scaled to the initial temperature as $\Delta ^{\ast
}=\Delta /k_{B}T(0)$. Time will be reported in units $\tau =\sqrt{\frac{%
m\sigma ^{2}}{2k_{B}T(0)}}$ and all calculations are performed with $p=1$ and using
the pressure given in eq.(\ref{pressure}) and the transport coefficients evaluated using
the adiabatic approximation introduced above.

\begin{figure}[tbp]
\begin{center}
\resizebox{17.4cm}{!}{
{\includegraphics{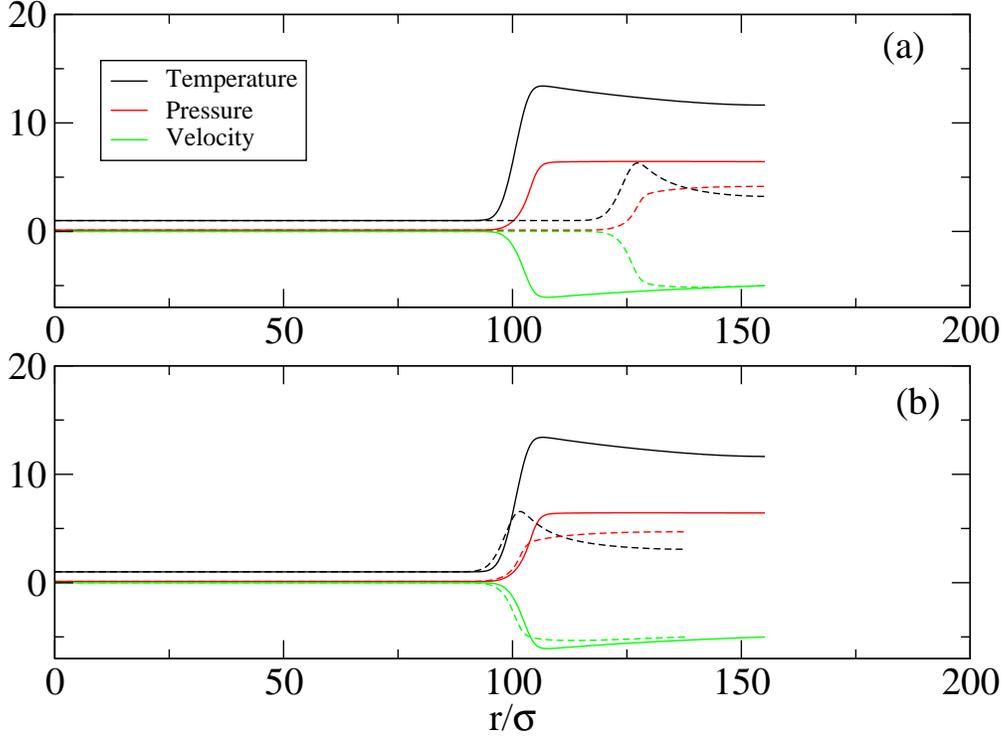}}}
\end{center}
\caption{Snapshots of the temperature, pressure and velocity profiles as
functions of distance from the center of the circle, $r$, for the elastic (full lines) and inelastic (broken
lines) gases. The upper lines are the temperature, the middle lines the
pressure and the lower lines the velocities.Figure (a) shows profiles at
equal times, $t=14.28\protect\tau$ while in Fig.(b) the inelastic profile
was taken at $t=17.79\protect\tau$ so as to show the shocks at approximately
equal positions.}
\label{fig6}
\end{figure}

\begin{figure}[tbp]
\begin{center}
\resizebox{17.4cm}{!}{
{\includegraphics{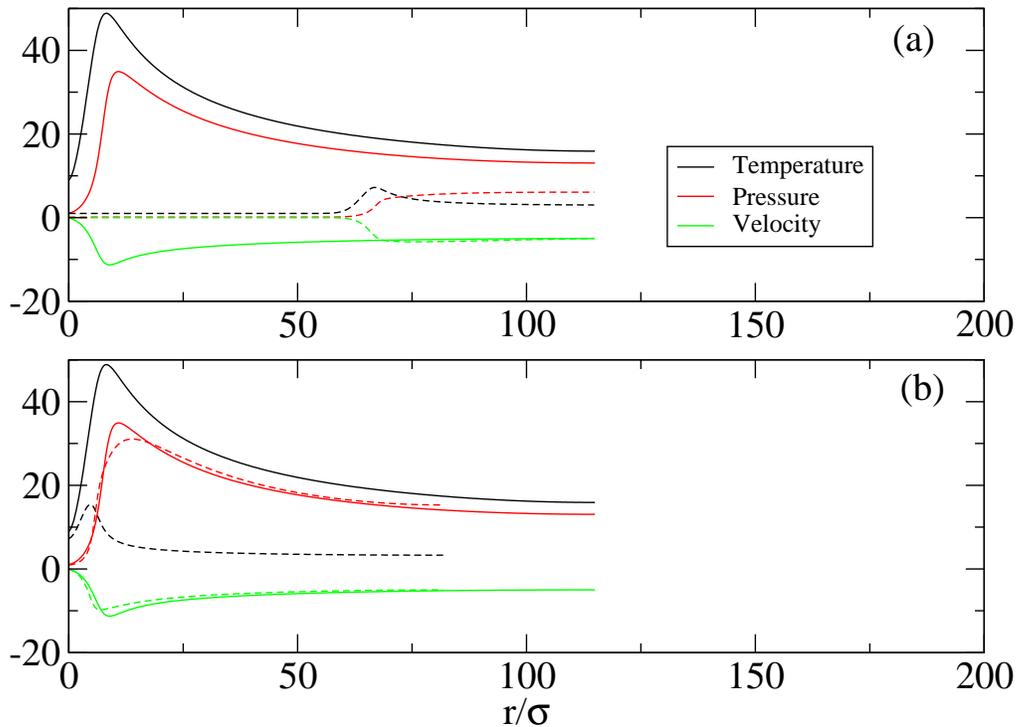}}}
\end{center}
\caption{The same as Fig. \ref{fig6} except for times when the shocks have
nearly reached the center of the bubbles. The times shown are $t=22.3\protect%
\tau$ in Fig. (a) and for the elastic profile in Fig.(b) and $t=28.96\protect%
\tau$ for the inelastic profile in Fig. (b).}
\label{fig7}
\end{figure}

\begin{figure}[tbp]
\begin{center}
\resizebox{17.4cm}{!}{
{\includegraphics{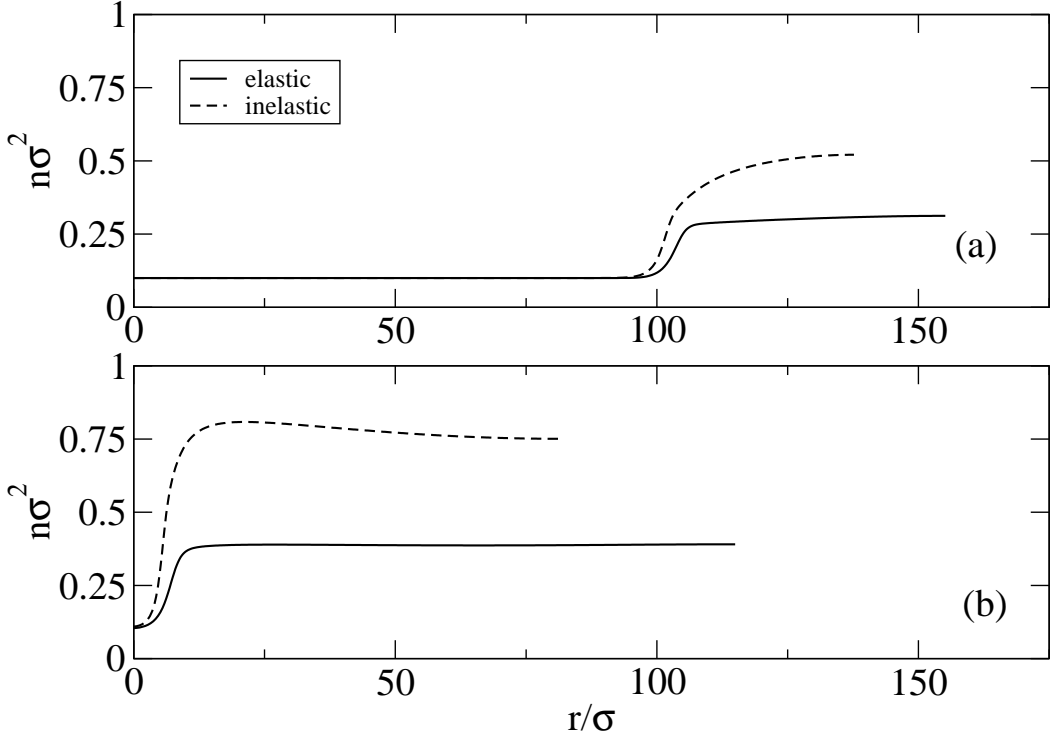}}}
\end{center}
\caption{Density profiles for the early (a) and later (b) times of figs. \ref%
{fig6}(b) and \ref{fig7}(b) respectively. The full lines are for the elastic gas
and the broken lines for the inelastic gas.}
\label{fig8}
\end{figure}

Figure \ref{fig6}(a) shows a comparison of the resulting shock waves for the
two cases $\Delta ^{\ast }=0$ , an elastic gas, and $\Delta ^{\ast }=20$, a
highly inelastic gas, after the same elapsed time after the shock profiles
are well developed. It is clear that the inelasticity leads to a substantial
slowing of the shock waves as well as a dramatic change in the shape of the
temperature profile. Shown in Fig. \ref{fig6}(b) is a comparison of the same
profiles for the elastic case to the profiles for the inelastic gas at a
later time when the shock waves are at approximately the same position. From
this comparison, it is clear that the pressure and velocity profiles are in
fact quite similar for the two cases and that most difference lies in the
temperature profile where that of the inelastic gas appears more like a
localized pulse than a shock. A similar comparison is made in Fig. \ref{fig7}
for a time just before the shocks are focused at the center of the bubble.
The same qualitative features appear:\ the shock in the inelastic gas is
significantly retarded compared to the elastic gas and all profiles are
similar, when compared at equal shock position, except for the temperature
profile. The fact that the pressure profiles are similar even though the
temperature profiles are dramatically different is due to a compensation in
the density profiles as shown in Fig. \ref{fig8}. Since the shocks in the
inelastic gas travel much more slowly than those in the elastic gas, the
density in the high-density region is necessarily higher in the inelastic
case since mass must be conserved. At the high densities that occur as the
shocks are focused, the excluded volume effects become significant and lead
to large contributions to the pressure. In fact, the densities are so high,
close to random close packing, that the model used for the pair distribution
function at contact, $\chi \left( n\right) $, is inaccurate and the pressure
effects are undoubtedly underestimated.

\begin{figure}[tbp]
\begin{center}
\resizebox{17.4cm}{!}{
{\includegraphics{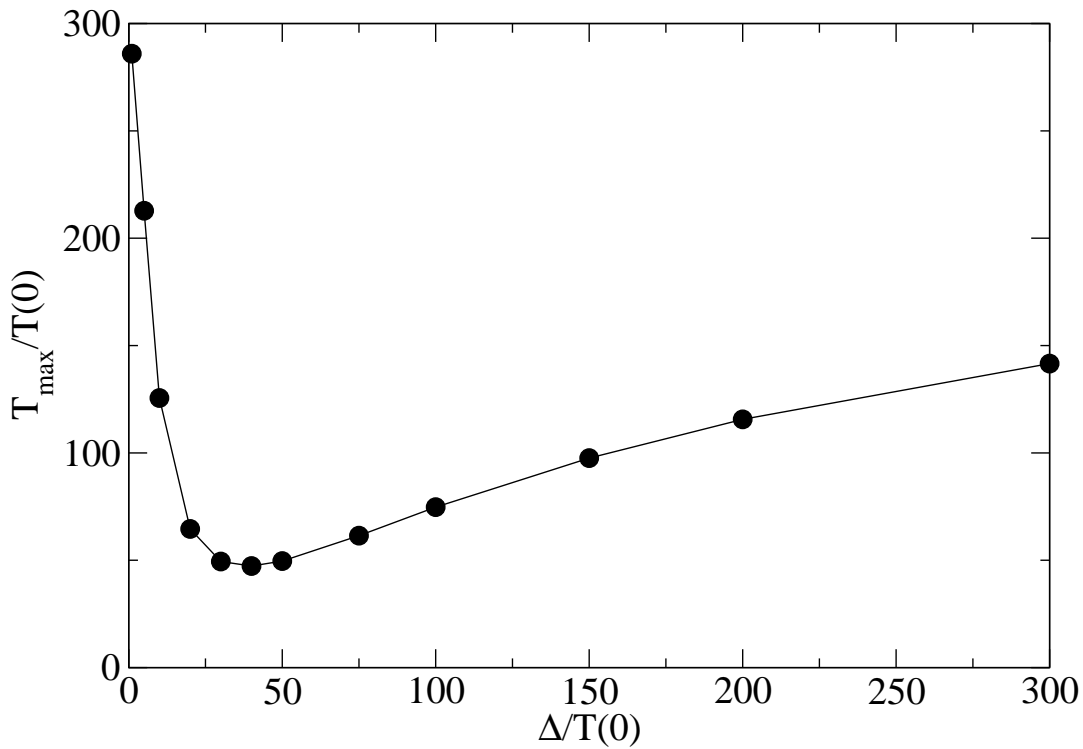}}}
\end{center}
\caption{Maximum temperature as a function of $\Delta$.}
\label{fig9}
\end{figure}

The maximum temperature achieved in the center of the bubble, $T_{\max },$
is shown in Fig. \ref{fig9} as a function of $\Delta ^{\ast }$. Starting
from the elastic limit, $\lim_{\Delta ^{\ast }\rightarrow \infty }T_{\max
}=198T(0)$, the maximum temperature decreases as $\Delta ^{\ast }$ decreases
as would be expected since lower values mean that more atoms can participate
in inelastic collisions. A minimum is reached near $\Delta ^{\ast }=40$
where $T_{\max }\sim 47T(0)$. It is interesting that at this minimum in $%
T_{\max }\left( \Delta ^{\ast }\right) $, one has $\Delta /T_{\max }\sim 1$
so that the cooling that occurs is maximal according to eq.(\ref{temp1}).
For smaller values of $\Delta ^{\ast }$, the maximum temperature increases
until at $\Delta ^{\ast }=1$ the maximum temperature is $284T(0)$ or about $%
50\%$ \emph{higher} than in the elastic, $\Delta ^{\ast }=0$, case. This is
due to the fact that as $\Delta ^{\ast }$ is lowered, the speed of the
shocks continues to decrease and the density behind the shocks continues to
increase leading to large pressure gradients when the shock focuses which
leads to the elevated temperatures. This interpretation is supported by the
fact that eliminating the divergence of the pressure at high densities by
setting $\chi \left( n\right) \rightarrow \chi \left( n(0)\right) $ in the
evaluation of the pressure eliminates the increase in maximal temperature as 
$\Delta ^{\ast }$ decreases to one. That only leaves the question as to why
the shock slows down. To answer that, the calculations were repeated with
all transport properties and the pressure are evaluated using the
expressions for an elastic gas so that the only effect of the inelasticity
is through the cooling rate. The result is that the shocks are still slow
and that the qualitative behavior of the maximal temperature is unchanged
leading to the conclusion that the slowing of the shock is solely due to the 
$\xi _{0}$ term in the heat equation.

As $\Delta ^{\ast }$ decreases from one, the maximum temperature begins to
decrease. However, the numerical code used becomes unstable in the range $%
0.8>\Delta ^{\ast }>0.2$ so it is not possible to follow this behavior
further. The instability may be due to deficiencies in the code but a more
intriguing possibility is that this indicates a real hydrodynamic
instability due to a pressure inversion as occurs in granular fluids. A
pressure inversion occurs when the compressibility becomes negative due to
the fact that an increase in density leads to such a decrease in temperature
that the pressure decreases with increasing density.

\section{Conclusions}

In this paper, a simple model of a gas exhibiting couplings between
hydrodynamics and chemistry, in the form of inelastic collisions, has been
explored. The model consists of hard-spheres that lose a fixed amount of
energy, $\Delta $, if the collisional energy is sufficient. This model was
inspired by the simplest prototype for an activated chemical reaction
wherein the reaction can only take place if there is sufficient kinetic
energy during collisions\cite{McQuarry}. It was found that in contrast to
the well known behavior of granular systems, the fixed energy loss model
studied here gives rise to a homogeneous cooling state that is well
described by a Gaussian distribution. The cooling is algebraic in time for
temperatures much larger than the energy loss, $\Delta $, and logarithmic
for temperatures much smaller than $\Delta $. While the cooling rate grows
monotonically with temperature, it was found that the rate of change of the
physically relevant variable $\Delta ^{\ast }=\Delta /k_{B}T$ has a maximum
at the value $\Delta ^{\ast }=1.5$ and that this gives rise to numerous
non-monotonic temperature dependencies in the thermodynamic and transport
properties of the system. In particular in the homogeneous state, the
coefficient of the first non-Gaussian corrections to the distribution
involve rapid changes around this value and yet are well predicted by
kinetic theory as demonstrated by comparison to numerical solutions of the
Enskog equation. Explicit expressions for the transport coefficients were
given and it was shown that a relatively simple and accurate ''adiabatic''\
approximation existed. The transport coefficients show ''resonances'' which
are again related to the non-monotonicity of the rate of change of $\Delta
^{\ast }$.

As an application, the hydrodynamic description thus specified was used to
study the behavior of shocks in a two-dimensional circular piston. It was
found that the maximum temperature obtained when the shocks were focussed at
the center of the volume exhibits a minimum as a function of $\Delta
/k_{B}T(0)$ at about $\Delta /k_{B}T(0)=40$ and rises to a maximum near $%
\Delta /k_{B}T(0)=1$. Below this value it appears that the fluid becomes
unstable, perhaps due to a pressure inversion. It is interesting to note
that for a gas at room temperature, $k_{B}T(0)\sim 300K$ so the minimum
occurs at $\Delta \sim 12,000K$ or, roughly, 1eV which is clearly a
physically relevant value for sonochemistry. The picture here is therefore
more complex than that found in the study of the affect of energy loss on
the maximum temperature using adiabatic models (i.e., assuming uniform
density and temperature in the gas)\cite{SonoEnergyLoss},\cite%
{SonoEnergyLossChallenged}. On the one hand, the present results show that
for a large range of values, $\Delta >12,000K$, the dissipation leads to a
substantial reduction of the maximum temperature - at the minimum, the
maximum temperature is only $20\%$ of that predicted for an elastic system -
in agreement with the predictions of Yasui\cite{SonoEnergyLoss}. On the
other hand, for the physically important range $300K<\Delta <12,000K$ the
maximal temperature increases rapidly with decreasing $\Delta $, eventually
reaching values $50\%$ greater than those predicted in an elastic system.
This increase is attributed entirely to excluded volume effects, in
agreement with the results of Toefgel, et al.\cite{SonoEnergyLossChallenged}%
, driven by the effect of the cooling in slowing down the shocks and leading
to much increased densities behind the shock front.

The present study is only a first step in understanding the coupling of
chemistry and hydrodynamics that is undoubtedly important in sonochemistry.
Several open questions remain. First, it would be useful to be able to
relate the cooling to the shock velocity. In particular, one could imagine
doing this at the level of the Euler equations, which are analytically
tractable in the case of an elastic fluid\cite{Gaspard}. Second, it would be
interesting to consider richer models with multiple species and real
endothermic chemical interactions. Finally, and perhaps most interesting,
would be to simply extend these results to 3 dimensions and with a coupling
to the Rayleigh-Plesset equation.

\begin{acknowledgments}
This work was supported in part by the European Space Agency under contract
number C90105.
\end{acknowledgments}

\appendix

\section{Details of the HCS calculations}

\label{AppA}

In ref.\cite{LutskoCE}, it was shown that the coefficients $I_{rs,k}$ which
determine the HCS could be calculated as%
\begin{equation}
I_{rs,k}=-n^{\ast }\frac{\Gamma \left( \frac{1}{2}D\right) \Gamma \left(
k+1\right) }{\Gamma \left( \frac{1}{2}D+k\right) }\left( \frac{2k_{B}T}{%
m\sigma ^{2}}\right) ^{1/2}\frac{1}{r!s!k!}\lim_{z_{1}\rightarrow
0}\lim_{z_{2}\rightarrow 0}\lim_{x\rightarrow 0}\frac{\partial ^{r}}{%
\partial z_{1}^{r}}\frac{\partial ^{s}}{\partial z_{2}^{s}}\frac{\partial
^{k}}{\partial x^{k}}\left[ \sum_{a}G_{a}\left( \Delta _{a}\right) -G_{0}%
\right]
\end{equation}%
with

\begin{eqnarray}
G_{a}\left( \psi _{t}|\Delta _{a}\right)  &=&-\frac{1}{2}\pi
^{-1/2}S_{D}\left( 1-z_{1}x\right) ^{-\frac{1}{2}D}\allowbreak \left( \frac{%
1-z_{1}x}{2-x-z_{2}-z_{1}+xz_{1}z_{2}}\right) ^{\frac{1}{2}} \\
&&\times \int_{0}^{\infty }du\;K_{a}^{\ast }\left( \sqrt{u}\right) \exp
\left( \frac{\left( 2-z_{2}-z_{1}\right) x}{2-x-z_{2}-z_{1}+xz_{1}z_{2}}%
\frac{1}{2}\Delta _{a}^{\ast }\left( \sqrt{u}\right) \right)   \notag \\
&&\times \exp \left( -\frac{1-z_{2}x}{2-x-z_{2}-z_{1}+xz_{1}z_{2}}u\right)  
\notag \\
&&\times \exp \left( -\frac{1}{2}\frac{\left( z_{2}-z_{1}\right) x}{%
2-x-z_{2}-z_{1}+xz_{1}z_{2}}\left( u-\sqrt{u}\sqrt{u-2\Delta _{a}^{\ast
}\left( \sqrt{u}\right) }\right) \right)   \notag
\end{eqnarray}%
and%
\begin{equation}
G_{0}=-\frac{1}{2}\pi ^{-1/2}S_{D}\left( 1-z_{1}x\right) ^{-\frac{D+1}{2}%
}\left( 2-x-z_{2}-z_{1}+xz_{1}z_{2}\right) ^{\frac{1}{2}}.
\end{equation}%
In the present case, 
\begin{eqnarray}
K_{0}^{\ast }\left( \sqrt{u}\right)  &=&K_{0}\left( \sqrt{\frac{2u}{m\beta }}%
\right) =1-p\Theta \left( \frac{u}{2\beta }-\Delta \right) =1-p\Theta \left(
u-2\beta \Delta \right)  \\
K_{1}^{\ast }\left( \sqrt{u}\right)  &=&K_{1}\left( \sqrt{\frac{2u}{m\beta }}%
\right) =p\Theta \left( \frac{u}{2\beta }-\Delta \right) =p\Theta \left(
u-2\beta \Delta \right)   \notag
\end{eqnarray}%
and $\Delta _{0}^{\ast }\left( \sqrt{u}\right) =0$ and $\Delta _{1}^{\ast
}\left( \sqrt{u}\right) =\beta \Delta $. For this model , the coefficients
can be more conveniently written as 
\begin{equation}
I_{rs,k}=I_{rs,k}^{E}+I_{rs,k}^{I}
\end{equation}%
with an elastic part defined by%
\begin{equation}
I_{rs,k}^{E}=-n^{\ast }\frac{\Gamma \left( \frac{1}{2}D\right) \Gamma \left(
k+1\right) }{\Gamma \left( \frac{1}{2}D+k\right) }\left( \frac{2k_{B}T}{%
m\sigma ^{2}}\right) ^{1/2}\frac{1}{r!s!k!}\lim_{z_{1}\rightarrow
0}\lim_{z_{2}\rightarrow 0}\lim_{x\rightarrow 0}\frac{\partial ^{r}}{%
\partial z_{1}^{r}}\frac{\partial ^{s}}{\partial z_{2}^{s}}\frac{\partial
^{k}}{\partial x^{k}}\left[ G_{E}-G_{0}\right] 
\end{equation}%
where the elastic generating function is%
\begin{eqnarray}
G_{E} &=&-\frac{1}{2}\pi ^{-1/2}S_{D}\left( 1-z_{1}x\right) ^{-\frac{1}{2}%
D}\allowbreak \left( \frac{1-z_{1}x}{2-x-z_{2}-z_{1}+xz_{1}z_{2}}\right) ^{%
\frac{1}{2}} \\
&&\times \int_{0}^{\infty }du\;\exp \left( -\frac{1-z_{2}x}{%
2-x-z_{2}-z_{1}+xz_{1}z_{2}}u\right)   \notag \\
&=&-\frac{1}{2}\pi ^{-1/2}S_{D}\left( 1-z_{1}x\right) ^{-\frac{1}{2}%
D}\allowbreak \left( \frac{2-x-z_{2}-z_{1}+xz_{1}z_{2}}{1-z_{2}x}\right) ^{%
\frac{1}{2}}  \notag
\end{eqnarray}%
and an inelastic contribution%
\begin{equation}
I_{rs,k}^{I}=-pn^{\ast }\frac{\Gamma \left( \frac{1}{2}D\right) \Gamma
\left( k+1\right) }{\Gamma \left( \frac{1}{2}D+k\right) }\left( \frac{2k_{B}T%
}{m\sigma ^{2}}\right) ^{1/2}\frac{1}{r!s!k!}\lim_{z_{1}\rightarrow
0}\lim_{z_{2}\rightarrow 0}\lim_{x\rightarrow 0}\frac{\partial ^{r}}{%
\partial z_{1}^{r}}\frac{\partial ^{s}}{\partial z_{2}^{s}}\frac{\partial
^{k}}{\partial x^{k}}G\left( \Delta ^{\ast }\right) 
\end{equation}%
with the inelastic generating function%
\begin{eqnarray}
G\left( \Delta \right)  &=&-\frac{1}{2}\pi ^{-1/2}S_{D}\left(
1-z_{1}x\right) ^{-\frac{1}{2}D}\allowbreak \left( \frac{1-z_{1}x}{%
2-x-z_{2}-z_{1}+xz_{1}z_{2}}\right) ^{\frac{1}{2}} \\
&&\times \int_{2\Delta }^{\infty }du\;\exp \left( -\frac{1}{2}\frac{%
2-xz_{1}-xz_{2}}{2-x-z_{2}-z_{1}+xz_{1}z_{2}}u\right)   \notag \\
&&\times \left[ 
\begin{array}{c}
\exp \left( \frac{1}{2}\frac{\left( 2-z_{2}-z_{1}\right) x\Delta +\left(
z_{2}-z_{1}\right) x\sqrt{u}\sqrt{u-2\Delta }}{2-x-z_{2}-z_{1}+xz_{1}z_{2}}%
\right)  \\ 
-\exp \left( \frac{1}{2}\frac{\left( z_{2}-z_{1}\right) xu}{%
2-x-z_{2}-z_{1}+xz_{1}z_{2}}\right) 
\end{array}%
\right] .  \notag
\end{eqnarray}%
A change of variables in the integral gives

\begin{eqnarray}
G\left( \Delta \right)  &=&-\frac{1}{2}\pi ^{-1/2}S_{D}\left(
1-z_{1}x\right) ^{-\frac{1}{2}D}\allowbreak \left( \frac{1-z_{1}x}{%
2-x-z_{2}-z_{1}+xz_{1}z_{2}}\right) ^{\frac{1}{2}} \\
&&\times \exp \left( -\frac{1}{2}\frac{2-xz_{1}-xz_{2}}{%
2-x-z_{2}-z_{1}+xz_{1}z_{2}}\Delta \right)   \notag \\
&&\times \Delta \int_{1}^{\infty }\exp \left( -\frac{1}{2}\frac{%
2-xz_{1}-xz_{2}}{2-x-z_{2}-z_{1}+xz_{1}z_{2}}\Delta y\right)   \notag \\
&&\times \left[ 
\begin{array}{c}
\exp \left( \frac{1}{2}\frac{\left( 2-z_{2}-z_{1}\right) x\Delta +\left(
z_{2}-z_{1}\right) x\Delta \sqrt{y^{2}-1}}{2-x-z_{2}-z_{1}+xz_{1}z_{2}}%
\right)  \\ 
-\exp \left( \frac{1}{2}\frac{\left( z_{2}-z_{1}\right) x\Delta \left(
y+1\right) }{2-x-z_{2}-z_{1}+xz_{1}z_{2}}\right) 
\end{array}%
\right] dy  \notag
\end{eqnarray}%
Using%
\begin{eqnarray}
&&\int_{1}^{\infty }\exp \left( -\frac{1}{2}\frac{2-xz_{1}-xz_{2}}{%
2-x-z_{2}-z_{1}+xz_{1}z_{2}}\Delta y+\frac{1}{2}\frac{\left(
z_{2}-z_{1}\right) x\Delta \sqrt{y^{2}-1}}{2-x-z_{2}-z_{1}+xz_{1}z_{2}}%
\right) dy \\
&=&\sum_{n=0}^{\infty }\frac{1}{n!}\left( \frac{1}{2}\frac{\left(
z_{2}-z_{1}\right) x\Delta }{2-x-z_{2}-z_{1}+xz_{1}z_{2}}\right)
^{n}\int_{1}^{\infty }\exp \left( -\frac{1}{2}\frac{2-xz_{1}-xz_{2}}{%
2-x-z_{2}-z_{1}+xz_{1}z_{2}}\Delta y\right) \left( y^{2}-1\right) ^{n/2}dy 
\notag \\
&=&\frac{2}{\sqrt{\pi }}\sum_{n=0}^{\infty }\frac{\Gamma \left( \frac{1}{2}%
n+1\right) }{\Gamma \left( n+1\right) }\Delta ^{\frac{n-1}{2}}F_{n}\left(
x,z_{1},z_{2};\Delta \right)   \notag
\end{eqnarray}%
with%
\begin{eqnarray}
F_{n}\left( x,z_{1},z_{2};\Delta \right)  &=&\left( z_{2}-z_{1}\right)
^{n}x^{n}\left( 2-x-z_{2}-z_{1}+xz_{1}z_{2}\right) ^{-\frac{n-1}{2}} \\
&&\times \left( 2-xz_{1}-xz_{2}\right) ^{-\frac{n+1}{2}}K_{\frac{n+1}{2}%
}\left( \frac{1}{2}\frac{2-xz_{1}-xz_{2}}{2-x-z_{2}-z_{1}+xz_{1}z_{2}}\Delta
\right)   \notag
\end{eqnarray}%
and%
\begin{eqnarray}
&&\int_{1}^{\infty }\exp \left( -\frac{1}{2}\frac{2-xz_{1}-xz_{2}}{%
2-x-z_{2}-z_{1}+xz_{1}z_{2}}\Delta y\right) \exp \left( \frac{1}{2}\frac{%
\left( z_{2}-z_{1}\right) x\Delta \left( y+1\right) }{%
2-x-z_{2}-z_{1}+xz_{1}z_{2}}\right) dy \\
&=&\exp \left( \frac{1}{2}\frac{\left( z_{2}-z_{1}\right) x\Delta }{%
2-x-z_{2}-z_{1}+xz_{1}z_{2}}\right) \int_{1}^{\infty }\exp \left( -\frac{%
1-z_{2}x}{2-x-z_{2}-z_{1}+xz_{1}z_{2}}\Delta y\right) dy  \notag \\
&=&\exp \left( -\frac{1}{2}\frac{2-3xz_{2}+xz_{1}}{%
2-x-z_{2}-z_{1}+xz_{1}z_{2}}\Delta \right) \left( \frac{1-z_{2}x}{%
2-x-z_{2}-z_{1}+xz_{1}z_{2}}\Delta \right) ^{-1}  \notag
\end{eqnarray}%
gives%
\begin{eqnarray}
G\left( \Delta \right)  &=&-\frac{1}{2}\pi ^{-1/2}S_{D}\left(
1-z_{1}x\right) ^{-\frac{1}{2}D}\allowbreak \left( \frac{1-z_{1}x}{%
2-x-z_{2}-z_{1}+xz_{1}z_{2}}\right) ^{\frac{1}{2}} \\
&&\times \exp \left( -\frac{1-x}{2-x-z_{2}-z_{1}+xz_{1}z_{2}}\Delta \right) 
\notag \\
&&\times \left[ 
\begin{array}{c}
\frac{2}{\sqrt{\pi }}\sum_{n=0}^{\infty }\frac{\Gamma \left( \frac{1}{2}%
n+1\right) }{\Gamma \left( n+1\right) }\Delta ^{\frac{n+1}{2}}F_{n}\left(
x,z_{1},z_{2};\Delta \right)  \\ 
-\left( \frac{2-x-z_{2}-z_{1}+xz_{1}z_{2}}{1-z_{2}x}\right) \exp \left( -%
\frac{1}{2}\frac{2-4z_{2}x+2x}{2-x-z_{2}-z_{1}+xz_{1}z_{2}}\Delta \right) 
\end{array}%
\right]   \notag
\end{eqnarray}%
Notice that in order to evaluate $I_{rs,k}$, only the first $r+s$ terms of
the sum need be retained. Thus, for example, using $K_{1/2}\left( x\right)
=\allowbreak \sqrt{\frac{\pi }{2x}}e^{-x}$ one has that%
\begin{equation}
\lim_{z_{1}\rightarrow 0}\lim_{z_{2}\rightarrow 0}G\left( \Delta \right) =-%
\frac{1}{2}\pi ^{-1/2}S_{D}\allowbreak \left( 2-x\right) ^{1/2}\left(
e^{-\Delta }-e^{\frac{2\Delta }{x-2}}\right) 
\end{equation}%
and%
\begin{eqnarray}
I_{k,00} &=&I_{k,00}^{I}=-pn^{\ast }\chi \frac{\Gamma \left( \frac{1}{2}%
D\right) }{\Gamma \left( \frac{1}{2}D+k\right) }\left( \frac{2k_{B}T}{%
m\sigma ^{2}}\right) ^{1/2}\lim_{z_{1}\rightarrow 0}\lim_{z_{2}\rightarrow
0}\lim_{x\rightarrow 0}\frac{\partial ^{k}}{\partial x^{k}}G\left( \Delta
^{\ast }\right)  \\
&=&\frac{S_{D}pn^{\ast }\chi }{2\sqrt{\pi }}\frac{\Gamma \left( \frac{1}{2}%
D\right) }{\Gamma \left( \frac{1}{2}D+k\right) }\left( \frac{2k_{B}T}{%
m\sigma ^{2}}\right) ^{1/2}\lim_{x\rightarrow 0}\frac{\partial ^{k}}{%
\partial x^{k}}\left( 2-x\right) ^{1/2}\left( e^{-\Delta ^{\ast }}-e^{\frac{%
2\Delta ^{\ast }}{x-2}}\right)   \notag
\end{eqnarray}%
so that%
\begin{eqnarray}
I_{1,00} &=&\frac{S_{D}pn^{\ast }\chi }{D\sqrt{\pi }}\left( \frac{k_{B}T}{%
m\sigma ^{2}}\right) ^{1/2}\Delta ^{\ast }e^{-\Delta ^{\ast }}\allowbreak  \\
I_{2,00} &=&\frac{S_{D}pn^{\ast }\chi }{D\left( D+2\right) \sqrt{\pi }}%
\left( \frac{k_{B}T}{m\sigma ^{2}}\right) ^{1/2}\left( 1-\Delta ^{\ast
}\right) \Delta ^{\ast }e^{-\Delta ^{\ast }}.  \notag
\end{eqnarray}

\section{Cooling Rate}

\label{AppB}

The cooling rate is well described by the equation%
\begin{equation}
\frac{d}{dt^{\ast }}\Delta ^{\ast }=\Delta ^{\ast 3/2}e^{-\Delta ^{\ast }}.
\label{C1}
\end{equation}%
If the initial temperature is high, so that $\Delta ^{\ast }\left( 0\right)
\ll 1$ , then for short times%
\begin{equation}
\frac{d}{dt^{\ast }}\Delta ^{\ast }\simeq \Delta ^{\ast 3/2}
\end{equation}%
which easily solved to give%
\begin{equation}
\Delta ^{\ast }\left( t^{\ast }\right) \simeq \frac{\Delta ^{\ast }\left(
0\right) }{\left( 1-\frac{1}{2}\sqrt{\Delta ^{\ast }\left( 0\right) }t^{\ast
}\right) ^{2}}
\end{equation}%
$\allowbreak $ suggesting that $\Delta ^{\ast }\rightarrow \infty $ and $%
T\rightarrow 0$ at $t^{\ast }=\sqrt{\frac{4k_{B}T\left( 0\right) }{\Delta }}$
. This obviously doesn't happen and instead, the long-time behavior,
corresponding to $\Delta ^{\ast }>>1$ can be determined by writing eq.(\ref%
{C1}) as%
\begin{eqnarray}
t^{\ast } &=&\int_{\Delta ^{\ast }\left( 0\right) }^{\Delta ^{\ast
}}x^{-3/2}e^{x}dx \\
&=&2\frac{e^{\Delta ^{\ast }\left( 0\right) }}{\sqrt{\Delta ^{\ast }\left(
0\right) }}+2i\sqrt{\pi }\text{erf}\left( i\sqrt{\Delta ^{\ast }\left(
0\right) }\right) -2\frac{e^{\Delta ^{\ast }}}{\sqrt{\Delta ^{\ast }}}-2i%
\sqrt{\pi }\text{erf}\left( i\sqrt{\Delta ^{\ast }}\right)  \notag
\end{eqnarray}%
Now%
\begin{equation}
\text{erf}\left( ix\right) =i\pi ^{-1/2}e^{x^{2}}\left( \frac{1}{x}+\frac{1}{%
2x^{3}}+\frac{3}{4x^{5}}+...\right)
\end{equation}%
so, defining $\gamma =-2\frac{e^{\Delta ^{\ast }\left( 0\right) }}{\sqrt{%
\Delta ^{\ast }\left( 0\right) }}-2i\sqrt{\pi }\text{erf}\left( i\sqrt{%
\Delta ^{\ast }\left( 0\right) }\right) $%
\begin{eqnarray}
t^{\ast } &=&-2\frac{e^{\Delta ^{\ast }}}{\sqrt{\Delta ^{\ast }}}+2\frac{%
e^{\Delta ^{\ast }}}{\sqrt{\Delta ^{\ast }}}\left( 1+\frac{1}{2\Delta ^{\ast
}}+\frac{3}{4\Delta ^{\ast 2}}+...\right) -\gamma \\
&=&\frac{e^{\Delta ^{\ast }}}{\Delta ^{\ast 3/2}}\left[ 1+\frac{3}{2\Delta
^{\ast }}+...-\gamma e^{-\Delta ^{\ast }}\Delta ^{\ast -3/2}\right]  \notag
\end{eqnarray}%
Taking only the leading order term, the solution is%
\begin{equation}
\Delta ^{\ast }=-\frac{3}{2}W\left( -\frac{2}{3t^{\ast \frac{2}{3}}}\right)
\end{equation}%
where $W(x)$ is the Lambert W-function\cite{AbramStegun}. For large
arguments, it behaves as $W\left( x\right) =\ln x-\ln \ln x+...$ so for
large times,%
\begin{equation}
\Delta ^{\ast }\simeq \ln \left( t\right)
\end{equation}

\section{The transport coefficients}

\label{App:Transport}

\subsection{Total transport coefficients}

With the approximation $c_{2}=0$, the transport coefficients can be written
in terms of their kinetic contributions as%
\begin{eqnarray}
\eta &=&\left( 1+\frac{2}{3}\overline{\theta }\right) \eta ^{K}+\frac{D}{D+2}%
\eta _{0}\overline{\gamma }_{1}  \label{c1} \\
\gamma &=&\gamma ^{K}+\eta _{0}\overline{\gamma }_{1}  \notag \\
\mu &=&\left( 1+\overline{\theta }\right) \mu ^{K}  \notag \\
\kappa &=&\left( 1+\overline{\theta }\right) \kappa ^{K}+\frac{D}{2}\frac{%
k_{B}}{m}\eta _{0}\overline{\gamma }_{1}  \notag
\end{eqnarray}%
with the auxiliary functions%
\begin{eqnarray}
\gamma _{1} &=&\eta _{0}\frac{4S_{D}^{2}}{\pi \left( D+2\right) D^{2}}%
n^{\ast 2}\chi \left[ 1+\frac{1}{4}p\int_{\sqrt{2\Delta ^{\ast }}}^{\infty
}e^{-\frac{1}{2}v^{2}}v^{2}\left( \sqrt{v^{2}-2\Delta ^{\ast }}-v\right) dv%
\right] \\
&=&\eta _{0}\frac{4S_{D}^{2}}{\pi \left( D+2\right) D^{2}}n^{\ast 2}\chi %
\left[ 1+\frac{1}{4}p\left( \Delta ^{\ast }e^{-\frac{1}{2}\Delta ^{\ast
}}K_{1}\left( \frac{1}{2}\Delta ^{\ast }\right) -2e^{-\Delta ^{\ast }}\left(
1+\Delta ^{\ast }\right) \right) \right]  \notag
\end{eqnarray}%
and%
\begin{eqnarray}
\overline{\theta } &=&\frac{3S_{D}}{2D\left( D+2\right) }n^{\ast }\chi \left[
1+\frac{1}{2\sqrt{2\pi }}p\int_{\sqrt{2\Delta ^{\ast }}}^{\infty }ve^{-\frac{%
1}{2}v^{2}}\left( v^{2}-1\right) \left( \sqrt{v^{2}-2\Delta ^{\ast }}%
-v\right) dv\right] \\
&=&\frac{3S_{D}}{2D\left( D+2\right) }n^{\ast }\chi \left[ 1+\frac{1}{2}%
p\left( e^{-\Delta }\left( 1+\Delta -2\sqrt{\frac{\Delta }{\pi }}\left(
1+\Delta \right) \right) -\left( 1-\text{erf}\left( \sqrt{\Delta }\right)
\right) \right) \right]  \notag
\end{eqnarray}%
It is also useful to note that%
\begin{eqnarray}
\frac{\partial \overline{\gamma }}{\partial \Delta ^{\ast }} &=&\frac{%
4S_{D}^{2}n^{\ast 2}\chi }{D^{2}\left( D+2\right) \pi }\left[ \frac{1}{4}%
p\left( -\frac{1}{2}\allowbreak e^{-\frac{1}{2}\Delta ^{\ast }}\Delta ^{\ast
}\left( K_{0}\left( \frac{1}{2}\Delta ^{\ast }\right) +K_{2}\left( \frac{1}{2%
}\Delta ^{\ast }\right) \right) +\allowbreak 2\Delta ^{\ast }e^{-\Delta
^{\ast }}\right) \right] \\
&&-\frac{1}{8}p\Delta e^{-\frac{1}{2}\Delta }\left( K_{1}\left( \frac{1}{2}%
\Delta \right) +K_{0}\left( \frac{1}{2}\Delta \right) -4e^{-\frac{1}{2}%
\Delta }\right) \allowbreak  \notag
\end{eqnarray}%
and%
\begin{equation}
\frac{\partial \theta }{\partial \Delta ^{\ast }}=\frac{3S_{D}}{2D\left(
D+2\right) }n^{\ast }\chi \left[ \frac{1}{2}pe^{-\Delta }\left( \left(
2\Delta -1\right) \sqrt{\frac{\Delta }{\pi }}-\Delta \right) \right]
\end{equation}

\subsection{ Dimensionless equations for kinetic parts of the transport
coefficients}

The kinematic contributions obey%
\begin{equation}
\xi _{0}\left( D+2\right) \frac{1}{m}T\frac{\partial \mu ^{K}}{\partial T}%
+I_{11}^{n}\mu ^{K}-\xi _{0}\left( D+2\right) \frac{1}{m}T\left( \frac{%
\partial \ln n\chi _{0}}{\partial n}\right) \kappa ^{K}=\left( nk_{B}T\left( 
\frac{k_{B}T}{m}\right) \frac{D+2}{2}\right) \Omega _{1}^{n}  \label{l1}
\end{equation}%
\begin{equation*}
\xi _{0}\left( D+2\right) \frac{1}{m}\left[ T\frac{\partial \kappa ^{K}}{%
\partial T}+\left( T\frac{\partial }{\partial T}\ln \xi _{0}\right) \kappa
^{K}\right] +I_{11}^{T}\kappa ^{K}=mn\left( \frac{k_{B}T}{m}\right) ^{2}%
\frac{D+2}{2}\left( \Omega _{1}^{T}+\frac{n}{T}\frac{k_{B}T}{m}\frac{D\left(
D+2\right) }{2}\right) 
\end{equation*}%
\begin{align}
\frac{1}{4}\xi _{0}\frac{D+2}{k_{B}T}\left[ T\frac{\partial }{\partial T}%
a_{2}^{\left( \nabla u\right) }+2a_{2}^{\left( \nabla u\right) }\right]
+I_{22}^{\nabla u}a_{2}^{\left( \nabla u\right) }& =\Omega _{2}^{\nabla u} 
\notag \\
\xi _{0}\frac{D+2}{m}\left( \frac{2k_{B}T}{m}\right) T\frac{\partial \eta
^{K}}{\partial T}+I_{00}^{\partial u}\eta ^{K}& =mn\left( \frac{k_{B}T}{m}%
\right) ^{2}\left[ n\left( \frac{k_{B}T}{m}\right) D\left( D+2\right) -2%
\sqrt{\frac{D}{D-1}}\Omega _{0}^{\partial u}\right]   \notag
\end{align}%
where $a_{2}^{\nabla u}=\sqrt{\frac{m\sigma ^{2}}{k_{B}T}}\overline{\gamma }%
^{K}$. Introducing the scaled transport coefficients and switching to $%
\Delta ^{\ast }$ as the independent variable gives%
\begin{eqnarray}
&&-\xi _{0}\left( D+2\right) \frac{1}{m}\kappa _{0}\frac{T}{n}\Delta ^{\ast }%
\frac{\partial \overline{\mu }^{K}}{\partial \Delta ^{\ast }}+\left[ \xi
_{0}\left( D+2\right) \frac{1}{m}\frac{3}{2}\kappa _{0}\frac{T}{n}%
+I_{11}^{n}\kappa _{0}\frac{T}{n}\right] \overline{\mu }^{K} \\
&&-\xi _{0}\left( D+2\right) \frac{1}{m}T\left( \frac{\partial \ln n\chi _{0}%
}{\partial n}\right) \kappa _{0}\overline{\kappa }^{K}  \notag \\
&=&\left( nk_{B}T\left( \frac{k_{B}T}{m}\right) \frac{D+2}{2}\right) \Omega
_{1}^{n}  \notag
\end{eqnarray}%
\begin{eqnarray*}
&&-\xi _{0}\left( D+2\right) \frac{1}{m}\kappa _{0}\Delta ^{\ast }\frac{%
\partial \overline{\kappa }^{K}}{\partial \Delta ^{\ast }}+\left[ \frac{D+2}{%
2m}\xi _{0}+\frac{D+2}{m}T\frac{\partial }{\partial T}\xi _{0}+I_{11}^{T}%
\right] \kappa _{0}\overline{\kappa }^{K} \\
&=&mn\left( \frac{k_{B}T}{m}\right) ^{2}\frac{D+2}{2}\left( \Omega _{1}^{T}+%
\frac{n}{T}\frac{k_{B}T}{m}\frac{D\left( D+2\right) }{2}\right) 
\end{eqnarray*}%
\begin{equation*}
-\frac{1}{4}\xi _{0}\frac{D+2}{k_{B}T}\left[ T\sqrt{\frac{m\sigma ^{2}}{%
k_{B}T}}\frac{\partial }{\partial T}\overline{\gamma }^{K}\right] +\left( 
\frac{1}{4}\xi _{0}\frac{D+2}{k_{B}T}\sqrt{\frac{m\sigma ^{2}}{k_{B}T}}%
+I_{22}^{\nabla u}\sqrt{\frac{m\sigma ^{2}}{k_{B}T}}\right) \overline{\gamma 
}^{K}=\Omega _{2}^{\nabla u}
\end{equation*}

\begin{gather*}
-\xi _{0}\frac{D+2}{m}\left( \frac{2k_{B}T}{m}\right) \eta _{0}\Delta ^{\ast
}\frac{\partial \overline{\eta }^{K}}{\partial \Delta ^{\ast }}+\left[ \xi
_{0}\frac{D+2}{m}\left( \frac{2k_{B}T}{m}\right) \frac{1}{2}%
+I_{00}^{\partial u}\right] \eta _{0}\overline{\eta }^{K} \\
=mn\left( \frac{k_{B}T}{m}\right) ^{2}\left[ n\left( \frac{k_{B}T}{m}\right)
D\left( D+2\right) -2\sqrt{\frac{D}{D-1}}\Omega _{0}^{\partial u}\right] 
\end{gather*}%
Next, introduce dimensionless forms for the Boltzmann integrals 
\begin{eqnarray}
\xi _{0} &=&-\frac{Dnk_{B}T}{2}n\sigma ^{D}\chi \frac{S_{D}}{2D\left(
D+2\right) \sqrt{\pi }}\left( \frac{k_{B}T}{m\sigma ^{2}}\right)
^{1/2}I_{1,00}^{\ast } \\
I_{11}^{n} &=&I_{11}^{T}=I_{11}^{TE}I_{T}^{\ast }  \notag \\
I_{22}^{\nabla u} &=&I_{22}^{\nabla uE}I_{\gamma }^{\ast }  \notag \\
I_{00}^{\partial u} &=&I_{00}^{\partial uE}I_{\eta }^{\ast }  \notag
\end{eqnarray}%
where the elastic contributions are%
\begin{eqnarray}
I_{11}^{nE} &=&I_{11}^{TE}=n^{2}\sigma ^{D-1}S_{D}\chi \left( \frac{k_{B}T}{m%
}\right) ^{3/2}\frac{2\left( D-1\right) }{\sqrt{\pi }}  \label{l3} \\
I_{22}^{\nabla uE} &=&n^{2}\sigma ^{D-1}S_{D}\chi \left( \frac{k_{B}T}{m}%
\right) ^{1/2}\frac{\left( D-1\right) }{2\sqrt{\pi }}  \notag \\
I_{00}^{\partial uE} &=&\chi n^{2}\sigma ^{D-1}S_{D}\left( \frac{k_{B}T}{m}%
\right) ^{5/2}\frac{4D}{\sqrt{\pi }}  \notag
\end{eqnarray}%
thus giving, 
\begin{gather}
I_{1,00}^{\ast }\Delta ^{\ast }\frac{\partial \overline{\mu }^{K}}{\partial
\Delta ^{\ast }}+\left[ 8\left( D-1\right) I_{T}^{\ast }-\frac{3}{2}%
I_{1,00}^{\ast }\right] \overline{\mu }^{K}+I_{1,00}^{\ast }\left(
1+n\partial _{n}\chi _{0}\right) \overline{\kappa }^{K}=\frac{16m\left(
D-1\right) }{k_{B}T\left( D+2\right) D\chi }\Omega _{1}^{n} \\
I_{1,00}^{\ast }\Delta ^{\ast }\frac{\partial \overline{\kappa }^{K}}{%
\partial \Delta ^{\ast }}+\left[ 8\left( D-1\right) I_{T}^{\ast
}-2I_{1,00}^{\ast }-T\frac{\partial }{\partial T}I_{1,00}^{\ast }\right] 
\overline{\kappa }^{K}=8\frac{D-1}{\chi }+16m\frac{D-1}{k_{B}\left(
D+2\right) n\chi D}\Omega _{1}^{T}  \notag \\
I_{1,00}^{\ast }\Delta ^{\ast }\frac{\partial }{\partial \Delta ^{\ast }}%
\overline{\gamma }^{K}+\left[ 8\left( D-1\right) I_{\gamma }^{\ast
}-I_{1,00}^{\ast }\right] \overline{\gamma }^{K}=\frac{16}{n^{2}\sigma
^{D}\chi S_{D}}\sqrt{\pi }\Omega _{2}^{\nabla u}  \notag \\
I_{1,00}^{\ast }\Delta ^{\ast }\frac{\partial \overline{\eta }^{K}}{\partial
\Delta ^{\ast }}+\left[ 8DI_{\eta }^{\ast }-\frac{1}{2}I_{1,00}^{\ast }%
\right] \overline{\eta }^{K}=\frac{8}{\chi }D-8\frac{m}{k_{B}Tn\chi \left(
D+2\right) }\left[ 2\sqrt{\frac{D}{D-1}}\Omega _{0}^{\partial u}\right]  
\notag
\end{gather}%
Finally, introduce scaled sources 
\begin{eqnarray}
\Omega _{0}^{\partial u} &=&\Omega _{0}^{\partial uE}\Omega _{\eta }^{\ast }
\\
\Omega _{2}^{\nabla u} &=&\chi n^{2}\sigma ^{D}S_{D}\Omega _{\gamma }^{\ast }
\notag \\
\Omega _{1}^{T} &=&\Omega _{1}^{TE}\Omega _{T}^{\ast }  \notag \\
\Omega _{1}^{n} &=&\frac{16m\left( D-1\right) }{k_{B}T\left( D+2\right)
D\chi }\Omega _{\mu }^{\ast }  \notag
\end{eqnarray}%
with the elastic contributions%
\begin{eqnarray}
\Omega _{1}^{TE} &=&n^{2}\sigma ^{D}\chi S_{D}\frac{k_{B}T}{m}\frac{1}{T}%
\frac{3}{4} \\
\Omega _{1}^{\partial uE} &=&-n^{2}\sigma ^{D}\chi S_{D}\frac{k_{B}T}{m}%
\frac{1}{2}\sqrt{\frac{D-1}{D}}  \notag
\end{eqnarray}%
so%
\begin{align}
I_{1,00}^{\ast }\Delta ^{\ast }\frac{\partial \overline{\mu }^{K}}{\partial
\Delta ^{\ast }}+\left[ 8\left( D-1\right) I_{T}^{\ast }-\frac{3}{2}%
I_{1,00}^{\ast }\right] \overline{\mu }^{K}+I_{1,00}^{\ast }\left(
1+n\partial _{n}\chi _{0}\right) \overline{\kappa }^{K}& =\Omega _{\mu
}^{\ast } \\
I_{1,00}^{\ast }\Delta ^{\ast }\frac{\partial \overline{\kappa }^{K}}{%
\partial \Delta ^{\ast }}+\left[ 8\left( D-1\right) I_{T}^{\ast
}-2I_{1,00}^{\ast }-T\frac{\partial }{\partial T}I_{1,00}^{\ast }\right] 
\overline{\kappa }^{K}& =8\frac{D-1}{\chi }+12\frac{\left( D-1\right) S_{D}}{%
D\left( D+2\right) }n^{\ast }\Omega _{T}^{\ast }  \notag \\
I_{1,00}^{\ast }\Delta ^{\ast }\frac{\partial }{\partial \Delta ^{\ast }}%
\overline{\gamma }^{K}+\left[ 8\left( D-1\right) I_{\gamma }^{\ast
}-I_{1,00}^{\ast }\right] \overline{\gamma }^{K}& =16\sqrt{\pi }\Omega
_{\gamma }^{\ast }  \notag \\
I_{1,00}^{\ast }\Delta ^{\ast }\frac{\partial \overline{\eta }^{K}}{\partial
\Delta ^{\ast }}+\left[ 8DI_{\eta }^{\ast }-\frac{1}{2}I_{1,00}^{\ast }%
\right] \overline{\eta }^{K}& =\frac{8}{\chi }D+\frac{8S_{D}}{D+2}n^{\ast
}\Omega _{\eta }^{\ast }  \notag
\end{align}

\subsection{Boltzmann Integrals}

Within the present approximations, ref. \cite{LutskoCE} gives the Boltzmann
integrals 
\begin{eqnarray}
I_{rs}^{\gamma } &=&I_{rs}^{\gamma E}\left[ 1+p\int_{\sqrt{2\Delta ^{\ast }}%
}^{\infty }e^{-\frac{1}{2}v^{2}}v\left( \Delta ^{\ast }S_{rs}^{\gamma
}\left( v\right) +\frac{1}{4}v\left( \sqrt{v^{2}-2\Delta }-v\right) \right)
dv\right] \\
&=&I_{rs}^{\gamma E}\left[ 1+p\Delta ^{\ast }\int_{\sqrt{2\Delta ^{\ast }}%
}^{\infty }e^{-\frac{1}{2}v^{2}}vS_{rs}^{\gamma }\left( v\right) dv-\frac{1}{%
2}pe^{-\Delta ^{\ast }}\left( 1+\Delta ^{\ast }\right) +\frac{1}{4}p\Delta
^{\ast }e^{-\frac{1}{2}\Delta ^{\ast }}K_{1}\left( \frac{1}{2}\Delta ^{\ast
}\right) \right]  \notag
\end{eqnarray}%
with%
\begin{eqnarray}
S_{11}^{n}\left( v\right) &=&S_{11}^{T}\left( v\right) =\frac{D+8}{16\left(
D-1\right) }\left( v^{2}-1\right)  \label{l4} \\
S_{22}^{\nabla u}\left( v\right) &=&\frac{1}{64\left( D-1\right) }\left(
v^{6}-9v^{4}+\left( 8D+49\right) \allowbreak v^{2}-37-8D\right)  \notag \\
&&-\frac{1}{64\left( D-1\right) }\left( v^{4}-6\allowbreak v^{2}+3\right)
\Delta ^{\ast }  \notag \\
S_{00}^{\partial u}\left( v\right) &=&\frac{1}{4D}\left( v^{2}-1\right) . 
\notag
\end{eqnarray}%
For the shear viscosity this gives%
\begin{eqnarray}
I_{\eta }^{\ast } &\equiv &\left( \chi n^{2}\sigma ^{D-1}S_{D}\left( \frac{%
k_{B}T}{m}\right) ^{5/2}\frac{4D}{\sqrt{\pi }}\right) ^{-1}I_{00}^{\partial
u} \\
&=&1+\frac{1}{4D}p\Delta ^{\ast }\int_{\sqrt{2\Delta ^{\ast }}}^{\infty }e^{-%
\frac{1}{2}v^{2}}v\left( v^{2}-1\right) dv-\frac{1}{2}pe^{-\Delta ^{\ast
}}\left( 1+\Delta ^{\ast }\right) +\frac{1}{4}p\Delta ^{\ast }e^{-\frac{1}{2}%
\Delta ^{\ast }}K_{1}\left( \frac{1}{2}\Delta ^{\ast }\right)  \notag \\
&=&1+\frac{1}{4D}pe^{-\Delta ^{\ast }}\left( 2\Delta ^{\ast 2}+\Delta ^{\ast
}\left( 1-2D\right) -2D\right) +\frac{1}{4}p\Delta ^{\ast }e^{-\frac{1}{2}%
\Delta ^{\ast }}K_{1}\left( \frac{1}{2}\Delta ^{\ast }\right) ;  \notag
\end{eqnarray}%
for the bulk viscosity%
\begin{eqnarray}
I_{\gamma }^{\ast } &\equiv &\left( n^{2}\sigma ^{D-1}S_{D}\chi \left( \frac{%
k_{B}T}{m}\right) ^{1/2}\frac{\left( D-1\right) }{2\sqrt{\pi }}\right)
^{-1}I_{22}^{\nabla u} \\
&=&%
\begin{array}{c}
1+\frac{1}{64\left( D-1\right) }p\Delta ^{\ast }\int_{\sqrt{2\Delta ^{\ast }}%
}^{\infty }e^{-\frac{1}{2}v^{2}}v\left( \left( v^{6}-9v^{4}+\left(
8D+49\right) \allowbreak v^{2}-37-8D\right) -\left( v^{4}-6\allowbreak
v^{2}+3\right) \Delta ^{\ast }\right) dv \\ 
-\frac{1}{2}pe^{-\Delta ^{\ast }}\left( 1+\Delta ^{\ast }\right) +\frac{1}{4}%
p\Delta ^{\ast }e^{-\frac{1}{2}\Delta ^{\ast }}K_{1}\left( \frac{1}{2}\Delta
^{\ast }\right)%
\end{array}
\notag \\
&=&1+\frac{1}{64\left( D-1\right) }pe^{-\Delta ^{\ast }}\left( 4\Delta
^{\ast 4}-8\Delta ^{\ast 3}+\left( 75+16D\right) \Delta ^{\ast 2}+\left(
69-24D\right) \Delta ^{\ast }-32\left( D-1\right) \right)  \notag \\
&&+\frac{1}{4}p\Delta ^{\ast }e^{-\frac{1}{2}\Delta ^{\ast }}K_{1}\left( 
\frac{1}{2}\Delta ^{\ast }\right) ;  \notag
\end{eqnarray}%
and for the density and thermal conductivity%
\begin{eqnarray}
I_{\kappa }^{\ast } &\equiv &\left( n^{2}\sigma ^{D-1}S_{D}\chi \left( \frac{%
k_{B}T}{m}\right) ^{3/2}\frac{2\left( D-1\right) }{\sqrt{\pi }}\right)
^{-1}I_{11}^{TE} \\
&=&1+\frac{D+8}{16\left( D-1\right) }p\Delta ^{\ast }\int_{\sqrt{2\Delta
^{\ast }}}^{\infty }e^{-\frac{1}{2}v^{2}}v\left( v^{2}-1\right) dv-\frac{1}{2%
}pe^{-\Delta ^{\ast }}\left( 1+\Delta ^{\ast }\right) +\frac{1}{4}p\Delta
^{\ast }e^{-\frac{1}{2}\Delta ^{\ast }}K_{1}\left( \frac{1}{2}\Delta ^{\ast
}\right)  \notag \\
&=&1+\frac{1}{16\left( D-1\right) }pe^{-\Delta ^{\ast }}\left( \left(
2D+16\right) \Delta ^{\ast 2}+\left( 16-7D\right) \Delta ^{\ast
}-8D+8\right) +\frac{1}{4}p\Delta ^{\ast }e^{-\frac{1}{2}\Delta ^{\ast
}}K_{1}\left( \frac{1}{2}\Delta ^{\ast }\right)  \notag
\end{eqnarray}

\subsection{Sources}

Reference \cite{LutskoCE} gives the sources as 
\begin{equation}
\Omega _{rs}^{\gamma }=\Omega _{rs}^{\gamma E}+\chi n^{2}\sigma ^{D}S_{D}%
\frac{1}{\sqrt{2\pi }}p\int_{\sqrt{2\Delta ^{\ast }}}^{\infty }e^{-\frac{1}{2%
}v^{2}}vT_{rs}^{\gamma }\left( v\right) dv
\end{equation}%
with 
\begin{eqnarray}
T_{11}^{n}\left( v\right) &=&\frac{1}{2}\frac{\partial \ln n^{2}\chi }{%
\partial n}\left( \frac{k_{B}T}{m}\right) \frac{1}{4}\left( \left(
v^{2}-3\right) \left( \sqrt{v^{2}-2\Delta ^{\ast }}-v\right) -2\Delta ^{\ast
}\left( \left( \sqrt{v^{2}-2\Delta ^{\ast }}-v\right) +v\right) \right)
\label{l7} \\
T_{11}^{T}\left( v\right) &=&\frac{k_{B}}{16m}\left( \left(
v^{4}-4v^{2}+9\right) \left( \sqrt{v^{2}-2\Delta ^{\ast }}-v\right) -2\Delta
^{\ast }\left( \left( v^{2}-1\right) \left( \sqrt{v^{2}-2\Delta ^{\ast }}%
-v\right) +v^{3}+5v\right) \right)  \notag \\
T_{22}^{\nabla u}\left( v\right) &=&\frac{1}{8D}\left( \left( 2\Delta ^{\ast
}+3-v^{2}\right) \left( \sqrt{v^{2}-2\Delta ^{\ast }}-v\right) +v\Delta
^{\ast }\left( v^{2}-1-\Delta ^{\ast }\right) \right)  \notag \\
T_{00}^{\partial u}\left( v\right) &=&\frac{1}{2}\sqrt{\frac{D-1}{D}}\left( 
\frac{k_{B}T}{m}\right) \left( v\Delta ^{\ast }-\left( \sqrt{v^{2}-2\Delta
^{\ast }}-v\right) \right)  \notag
\end{eqnarray}

So%
\begin{eqnarray}
\Omega _{1}^{n} &=&\chi n^{2}\sigma ^{D}S_{D}\frac{1}{\sqrt{2\pi }}p\frac{1}{%
2}\frac{\partial \ln n^{2}\chi }{\partial n}\left( \frac{k_{B}T}{m}\right) 
\\
&&\times \int_{\sqrt{2\Delta ^{\ast }}}^{\infty }e^{-\frac{1}{2}v^{2}}v\frac{%
1}{4}\left( \left( v^{2}-3\right) \left( \sqrt{v^{2}-2\Delta ^{\ast }}%
-v\right) -2\Delta ^{\ast }\left( \left( \sqrt{v^{2}-2\Delta ^{\ast }}%
-v\right) +v\right) \right) dv  \notag \\
&=&-\chi n^{2}\sigma ^{D}S_{D}\frac{1}{4\sqrt{\pi }}p\frac{\partial \ln
n^{2}\chi }{\partial n}\left( \frac{k_{B}T}{m}\right) e^{-\Delta ^{\ast
}}\Delta ^{\ast \frac{3}{2}}  \notag \\
&=&\frac{k_{B}T\left( D+2\right) D\chi }{16m\left( D-1\right) }\Omega _{\mu
}^{\ast }  \notag
\end{eqnarray}%
with%
\begin{equation}
\Omega _{\mu }^{\ast }=-\frac{4\left( D-1\right) S_{D}}{\sqrt{\pi }\left(
D+2\right) D}n^{\ast }p\left( 2+n\partial _{n}\ln \chi \right) e^{-\Delta
^{\ast }}\Delta ^{\ast \frac{3}{2}}
\end{equation}%
and%
\begin{eqnarray}
\Omega _{1}^{\nabla u} &=&\chi n^{2}\sigma ^{D}S_{D}\frac{1}{8D\sqrt{2\pi }}%
p\int_{\sqrt{2\Delta ^{\ast }}}^{\infty }e^{-\frac{1}{2}v^{2}}v\left( 
\begin{array}{c}
\left( 2\Delta ^{\ast }+3-v^{2}\right) \left( \sqrt{v^{2}-2\Delta ^{\ast }}%
-v\right)  \\ 
+v\Delta ^{\ast }\left( v^{2}-1-\Delta ^{\ast }\right) 
\end{array}%
\right) dv \\
&=&\chi n^{2}\sigma ^{D}S_{D}\Omega _{\gamma }^{\ast }\allowbreak .  \notag
\end{eqnarray}%
with%
\begin{equation}
\Omega _{\gamma }^{\ast }=\frac{1}{16\sqrt{\pi }}\frac{1}{D}p\Delta \left( 2%
\sqrt{\Delta }\left( 2+\Delta \right) e^{-\Delta }-\Delta \sqrt{\pi }\left(
1-\text{erf}\left( \sqrt{\Delta }\right) \right) \right) .
\end{equation}%
For the shear%
\begin{equation}
\Omega _{1}^{\partial u}=-n^{2}\sigma ^{D}\chi S_{D}\frac{k_{B}T}{m}\frac{1}{%
2}\sqrt{\frac{D-1}{D}}\Omega ^{\ast }
\end{equation}%
with%
\begin{eqnarray}
\Omega ^{\ast } &=&1-\frac{1}{\sqrt{2\pi }}p\int_{\sqrt{2\Delta ^{\ast }}%
}^{\infty }e^{-\frac{1}{2}v^{2}}v\left( v\Delta ^{\ast }-\left( \sqrt{%
v^{2}-2\Delta ^{\ast }}-v\right) \right) dv \\
&=&1-\frac{1}{\sqrt{2\pi }}p\left[ \sqrt{2\Delta }\left( 1+\Delta \right)
e^{-\Delta }-\sqrt{\frac{\pi }{2}}e^{-\Delta }+\sqrt{\frac{\pi }{2}}\left(
1+\Delta \right) \left( 1-\text{erf}\left( \sqrt{\Delta }\right) \right) %
\right] \allowbreak   \notag
\end{eqnarray}%
and for the temperature%
\begin{equation}
\Omega _{1}^{T}=n^{2}\sigma ^{D}\chi S_{D}\frac{k_{B}T}{m}\frac{1}{T}\frac{3%
}{4}\Omega ^{\ast }
\end{equation}%
with%
\begin{eqnarray}
\Omega ^{\ast } &=&1+\frac{1}{12}\frac{1}{\sqrt{2\pi }}p\int_{\sqrt{2\Delta
^{\ast }}}^{\infty }e^{-\frac{1}{2}v^{2}}v\left( 
\begin{array}{c}
\left( v^{4}-4v^{2}+9\right) \left( \sqrt{v^{2}-2\Delta ^{\ast }}-v\right) 
\\ 
-2\Delta ^{\ast }\left( \left( v^{2}-1\right) \left( \sqrt{v^{2}-2\Delta
^{\ast }}-v\right) +v^{3}+5v\right) 
\end{array}%
\right) dv \\
&=&1+\frac{1}{2}p\left[ \left( 1-\frac{1}{3\sqrt{\pi }}\left( \Delta
+2\right) \left( 2\Delta +3\right) \sqrt{\Delta }\right) e^{-\Delta }-\left(
1+\Delta \right) \left( 1-\text{erf}\left( \sqrt{\Delta }\right) \right) %
\right] .  \notag
\end{eqnarray}

\subsection{Derivatives}

To implement the numerical solution of the Navier-Stokes equations, it is
useful to know the derivatives of the Boltzmann integrals and the sources.
For the convenience these are recorded here for the former,%
\begin{eqnarray}
\frac{\partial }{\partial \Delta ^{\ast }}I_{\eta }^{\ast } &=&\frac{1}{4}Dp%
\frac{\partial }{\partial \Delta }\left( \frac{1}{D}e^{-\Delta }\left(
2\Delta ^{2}+\left( 1-2D\right) \Delta -2D\right) +\Delta e^{-\frac{1}{2}%
\Delta }K_{1}\left( \frac{1}{2}\Delta \right) \right)  \\
&=&\frac{1}{4}Dp\left( \frac{1}{D}e^{-\Delta }\left( -2\Delta ^{2}+\Delta
\left( 3+2D\right) +1\right) \allowbreak -\frac{1}{2}\Delta e^{-\frac{1}{2}%
\Delta }\left( K_{1}\left( \frac{1}{2}\Delta \right) +K_{0}\left( \frac{1}{2}%
\Delta \right) \right) \allowbreak \right)   \notag \\
\frac{\partial }{\partial \Delta ^{\ast }}I_{\kappa }^{\ast } &=&-\frac{1}{%
16\left( D-1\right) }pe^{-\Delta }\left( 2\Delta ^{2}\left( D+8\right)
-\Delta \left( 16+11D\right) -D-8\right) \allowbreak   \notag \\
&&-\frac{1}{8}p\Delta e^{-\frac{1}{2}\Delta }\left( D-1\right) \left(
K_{1}\left( \frac{1}{2}\Delta \right) +K_{0}\left( \frac{1}{2}\Delta \right)
\right) \allowbreak   \notag \\
\frac{\partial }{\partial \Delta ^{\ast }}I_{\gamma }^{\ast } &=&-\frac{1}{%
64\left( D-1\right) }pe^{-\Delta }\left( 4\Delta ^{4}-24\Delta ^{3}+\left(
99+16D\right) \Delta ^{2}-\left( 81+56D\right) \Delta -8D-37\right)
\allowbreak   \notag \\
&&-\frac{1}{8}p\Delta e^{-\frac{1}{2}\Delta }\left( K_{1}\left( \frac{1}{2}%
\Delta \right) +K_{0}\left( \frac{1}{2}\Delta \right) \right)   \notag
\end{eqnarray}

and for the sources

\begin{eqnarray}
\frac{\partial }{\partial \Delta ^{\ast }}\Omega _{\eta }^{\ast } &=&\frac{1%
}{2}p\left( \text{erf}\left( \sqrt{\Delta }\right) -1+\left( \frac{2}{\sqrt{%
\pi }}\Delta ^{\frac{3}{2}}-1\right) e^{-\Delta }\right) \\
\frac{\partial }{\partial \Delta ^{\ast }}\Omega _{\kappa }^{\ast } &=&\frac{%
1}{6}p\left( \frac{-3\sqrt{\Delta }+4\Delta ^{\frac{3}{2}}+4\Delta ^{\frac{5%
}{2}}-6\sqrt{\pi }}{2\sqrt{\pi }}e^{-\Delta }+3\left( \text{erf}\left( \sqrt{%
\Delta }\right) -1\right) \right)  \notag \\
\frac{\partial }{\partial \Delta ^{\ast }}\Omega _{\mu }^{\ast } &=&-p\frac{2%
}{\sqrt{\pi }}\frac{D-1}{\left( D+2\right) D}n^{\ast }S_{D}\left(
2+n\partial _{n}\ln \chi \right) \sqrt{\Delta ^{\ast }}\left( 3-2\Delta
^{\ast }\right) e^{-\Delta ^{\ast }}  \notag \\
\frac{\partial }{\partial \Delta ^{\ast }}\Omega _{\gamma }^{\ast } &=&\frac{%
1}{8}\frac{1}{D}p\left( \left( 3+\Delta -\Delta ^{2}\right) \sqrt{\frac{%
\Delta }{\pi }}e^{-\Delta }-\Delta \left( 1-\text{erf}\left( \sqrt{\Delta }%
\right) \right) \right) \allowbreak  \notag
\end{eqnarray}

\subsection{Cooling}

From ref. \cite{LutskoCE}, the additional cooling term in the Navier-Stokes
equations is 
\begin{eqnarray}
\left( \overrightarrow{\nabla }\cdot \overrightarrow{u}\right) \xi _{1}
&=&\xi _{0}\left[ f_{1}\right] +\xi _{1}\left[ f_{0}\right]  \label{h1} \\
\xi _{0}\left[ f_{1}\right] &=&-\left( \overrightarrow{\nabla }\cdot 
\overrightarrow{u}\right) a_{2}^{\nabla u}n^{2}\sigma ^{D}\chi S_{D}\left( 
\frac{k_{B}T}{m\sigma ^{2}}\right) ^{1/2}\frac{k_{B}T}{32\sqrt{\pi }}p\int_{%
\sqrt{2\Delta ^{\ast }}}^{\infty }ve^{-\frac{1}{2}v^{2}}\Delta \left(
v^{4}-6v^{2}+3\right) dv  \notag \\
\xi _{1}\left[ f_{0}\right] &=&\left( \overrightarrow{\nabla }\cdot 
\overrightarrow{u}\right) n^{2}\sigma ^{D}\chi S_{D}\frac{k_{B}T}{2\sqrt{%
2\pi }D}p\int_{\sqrt{2\Delta ^{\ast }}}^{\infty }\Delta v^{2}e^{-\frac{1}{2}%
v^{2}}dv  \notag
\end{eqnarray}%
giving%
\begin{equation}
\xi _{1}=-p\left( nk_{B}T\right) n^{\ast }\chi S_{D}\Delta \left[ \frac{1}{32%
\sqrt{\pi }}\left( 4\Delta ^{2}-4\Delta -1\right) e^{-\Delta }\overline{%
\gamma }^{K}+\frac{1}{4D}\left( 2\sqrt{\frac{\Delta }{\pi }}e^{-\Delta
}+\left( 1-\text{erf}\left( \sqrt{\Delta }\right) \right) \right) \right]
\end{equation}

\bibliographystyle{prsty}
\bibliography{physics}

\end{document}